\documentclass[letterpaper,twocolumn,10pt]{article}

\usepackage{usenix}
\usepackage[utf8]{inputenc}
\usepackage[T1]{fontenc}
\usepackage{graphics}
\usepackage{amsmath}
\usepackage{amssymb}
\usepackage{pifont}
\usepackage{amssymb} 
\usepackage{rotating}
\usepackage{multirow}
\usepackage{makecell}
\usepackage{comment}
\usepackage{changepage}
\usepackage{xspace}
\usepackage{tikz}
\usepackage{tabto}
\usepackage{acronym}
\usepackage{tabularx}
\usepackage{paralist}
\usepackage{listings}
\usepackage{xcolor}
\usepackage{colortbl}
\usepackage{xurl}
\usepackage{enumitem}
\usepackage{subcaption}
\usepackage{amsfonts}
\usepackage{booktabs}
\usepackage{cite}
\usepackage{hyperref}

\newcommand{\cf}{cf.\@\xspace}
\newcommand{\ie}{i.\@\,e.\@\xspace}
\newcommand{\eg}{e.\@\,g.\@\xspace}

\newcommand{\etc}{etc.\@\xspace}
\newcommand{\etal}{et~al.\@\xspace}

\definecolor{dkgreen}{rgb}{0,0.6,0}
\definecolor{gray}{rgb}{0.5,0.5,0.5}
\definecolor{mauve}{rgb}{0.58,0,0.82}

\def\BibTeX{{\rm B\kern-.05em{\sc i\kern-.025em b}\kern-.08em
    T\kern-.1667em\lower.7ex\hbox{E}\kern-.125emX}}

\begin{document}

\title{Freaky Leaky SMS: Extracting User Locations by Analyzing SMS Timings}

\author{
{\rm Evangelos Bitsikas}\\
bitsikas.e@northeastern.edu\\
Northeastern University
\and
{\rm Theodor Schnitzler}\\
theodor.schnitzler@tu-dortmund.de\\
Research Center Trustworthy\\Data Science and Security
\and
{\rm Christina Pöpper}\\
christina.poepper@nyu.edu \\
New York University Abu Dhabi
\and
{\rm Aanjhan Ranganathan}\\
aanjhan@northeastern.edu\\
Northeastern University
}

\maketitle

\begin{abstract}

Short Message Service (SMS) remains one of the most popular communication channels since its introduction in 2G cellular networks. In this paper, we demonstrate that merely receiving silent SMS messages regularly opens a stealthy side-channel that allows other regular network users to infer the whereabouts of the SMS recipient. The core idea is that receiving an SMS inevitably generates Delivery Reports whose reception bestows a timing attack vector at the sender. We conducted experiments across various countries, operators, and devices to show that an attacker can deduce the location of an SMS recipient by analyzing timing measurements from typical receiver locations. Our results show that, after training an ML model, the SMS sender can accurately determine multiple locations of the recipient. For example, our model achieves up to 96\% accuracy for locations across different countries, and 86\% for two locations within Belgium. Due to the way cellular networks are designed, it is difficult to prevent Delivery Reports from being returned to the originator making it challenging to thwart this covert attack without making fundamental changes to the network architecture.

\end{abstract}

\section{Introduction}
Despite the emergence of smartphones and various messaging applications, the Short Message Service (SMS) remains an essential communication channel for sending and receiving text messages.
SMS is widely used in marketing campaigns, appointment reminders, short customer surveys, and even as part of two-factor authentication~\cite{peetersPMOST22}, identity verification, and security/emergency alerts~\cite{reavesSTBTB16,ReavesVSTBTB18}. 
Since its introduction in the GSM standard in the early 1990s, SMS remains a key service across cellular generations, including 5G standards~\cite{3gpp.23.501}. 

SMS's prevalence, global reach, and message delivery reliability have made it a significant attack vector in recent years. 
For example, smishing attacks~\cite{smishing-attack} use an SMS with malicious links to direct victims to a phishing website and deceive them into divulging sensitive information. 
The Flubot virus~\cite{flubot-attack} in 2021/22 spread via SMS containing links to trojan apps that accessed sensitive data like banking credentials, contacts, and disabled security options.
SMSes have also been used for spamming~\cite{sms-spam-attack}. Simjacker~\cite{adaptivemobile19:simjacking} and its variant WIBAttack~\cite{WIBattack-attack} are other malware examples that use binary-embedded SMS messages. 

In this work, we take an orthogonal approach and show how an attacker can subtly exploit SMS and determine a victim's location. 
Prior location identification and localization techniques in the cellular domain relied on retrieving the temporary and permanent identifiers of a mobile device using false base stations~\cite{Kotuliak22:l-track, Shaik2016:Practical-Attacks-LTE}, using them to track the user's whereabouts within a certain area. 
Authorities worldwide have used silent\footnote{Silent SMSes are not displayed by the victim's mobile.} SMSes~\cite{Matthias19:silent-sms-authorities} to uncover their owners' locations, however, these approaches rely on cooperation from the network operators and/or necessitate sniffers. 

Unlike the above attacks, our attack does not require access to the network operator's infrastructure or false base stations around the victim's area of interest. 
Instead, it works by leveraging SMS Delivery Reports, which are transmitted back to the sender when the network delivers the SMS to the recipient. 
The sender can request these reports, and there is no way for the recipient to prevent them. 
By measuring the round-trip time, \ie, the time elapsed between sending an SMS and receiving the corresponding Delivery Report, our attack can distinguish various locations of the target recipient and determine their location area after a training phase. 
The attacker can behave like a regular user and does not require access to advanced equipment, but a typical smartphone device.

Consider the following scenario of a nation's diplomat (victim) giving a press conference from a specific location, \eg, official residence. Given the public knowledge of the victim's current location and phone number, the adversary starts sending silent or regular SMSes to the victim and collects their round-trip time measurements, generating \emph{timing signatures} for that specific location. Then, at a later time, when the attacker wants to infer whether the victim is back in their residence, the adversary simply sends a silent SMS and determines whether the timing signatures match.  
Since Delivery Reports are solidly rooted in protocol specifications across all mobile network generations with no possibility to disable them, the attacker can reach the victim at any time by solely possessing their mobile phone number. It is hard to disable the attack without a significant overhaul of the cellular network specifications.

To the best of our knowledge, this work is the first to identify an SMS Delivery Report-based timing side-channel that leaks location information. Our work makes the following contributions: 

\begin{compactenum}
\item We enumerate the various cellular network components that contribute to the timing delays and identify six timing-related features to create a robust location signature. Based on the location signature, we design an approach to execute our SMS location inference attack.
\item We perform a large-scale study collecting Delivery Report timing measurements across three continents, nine countries, and ten operators to create our training dataset. We send SMS messages between devices and measure Delivery Reports return times within and across different setups in the US, multiple countries in Europe, and the Middle East. 
\item We use the collected measurements to evaluate the performance of our location inference attack. Our experiments show that we can achieve up to 75\% and 96\% accuracy for location identification in nearby and far countries, respectively. Our model achieves over 70\% for many cases within a country or certain region, such as within Germany, Netherlands and Belgium. We analyze factors affecting the accuracy of our location classification and perform network and temporal stability analyses for additional evaluation. 
\item We discuss potential countermeasures against SMS timing attacks, including enforcing random or uniform delays in the core network.
\end{compactenum}

In summary, we highlight the effectiveness of inferring location  based on SMS Delivery Reports and the challenges associated with mitigating such an attack. 

\noindent \textbf{Responsible Disclosure:} The privacy issues caused by the SMS timing attack have been recognized by GSMA on the \textit{GSMA Mobile Security Research Acknowledgements} page under \emph{CVD-2023-0072}. GSMA has been considering several countermeasures, including  artificial delays and robust SMS filtering. 


\noindent \textbf{Code Release:} The code along with the dataset are publicly available on Github at \url{https://github.com/vaggelis-sudo/SMS-Location-Identification-Attack}.

\section{Background on SMS Networks}

First, we explain the various types of network architectures that are used for SMS exchanges. Then, we illustrate how the SMS procedure in such networks works and which timing delays are involved, respectively.

\subsection{Cellular network architectures} \label{architectures}

\begin{figure*}[tb]
    \centering
    \includegraphics[width=1.\textwidth,keepaspectratio]{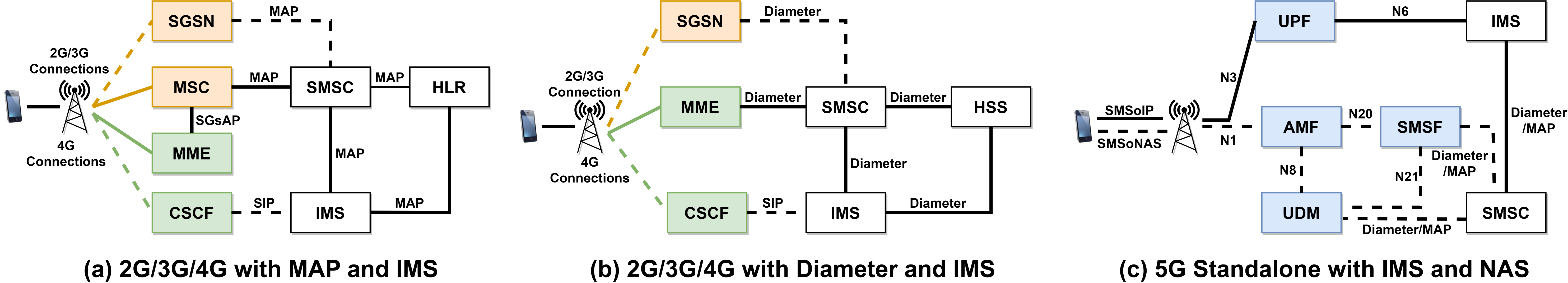}
    \caption{The SMS architectures based on the protocols and generations.}
    \label{fig:arch}
\end{figure*}

Figure~\ref{fig:arch} shows an extended version of 4G/LTE (a)--(b) and 5G standalone (c) architectures for the SMS procedure including the 2G/3G structures. In this work, we focus on LTE and 5G networks.

5G has two SMS delivery routing paths and protocols: SMSoIP and SMSoNAS. SMSoIP or IP-based communication (data-plane) leverages the SIP protocol and the IP Multimedia Subsystem (IMS) architecture~\cite{3gpp.33.203, 3gpp.24.341, 3gpp.23.228} to communicate with the Short Message Service Center (SMSC). SMSoNAS uses the Non-Access Stratum (NAS) protocol for SMS transmission and delivery, providing NAS encryption and integrity-protection~\cite{3gpp.33.501, 3gpp.24.501} through control-plane traffic after establishing the security context~\cite{3gpp.24.301, 3gpp.23.501}.

Furthermore, LTE services support chiefly IP-based communication through the IMS (Figure~\ref{fig:arch}), then alternatively the SGsAP interface, which eliminates the need for 2G/3G fallback, and finally, the NAS signaling communication combined with the Diameter protocol~\cite{ng.111-v2.0}. Typically, the IMS incorporates the IP Short Message Gateway (IP-SM-GW), an IMS Application Server that handles SIP-based messaging services for IMS subscribers.

The selection between SMSoNAS and SMSoIP depends on the SMS originator and the network support, even though IP-based communications are more prevalent, as the User Equipment (UE) subscribes to the IMS after completing the Authentication and Key Agreement (AKA) procedure with the Core Network. 


\subsection{SMS procedure}\label{sms-procedure}

SMS services are accessible to all network generations (2G-5G) \cite{ng.111-v2.0} as a process of exchanging short text messages between two network subscribers. The SMS exchange between originator and recipient requires forwarding to the Core Network, where the SMSC manages the SMS process and delivery (Figures~\ref{fig:sms-burst} and~\ref{fig:timings}). 
After receiving the message, the recipient sends a Delivery Report, which is forwarded through the SMSC to the originator acknowledging the delivery.
Delivery Reports provide detailed information on the status of every message sent including "Delivered", "Accepted", "Failed", "Undeliverable", "Expired", and "Rejected". 
Failed deliveries could be due to incorrect phone number, disabled international roaming, unreachable recipient, mobile plan restrictions, \etc.
Note that delivery notification is enabled by the originator in their phone's settings on modern smartphones.
Delivery Reports are used for data cleansing/updating, improving response rates, audit trail, and systems monitoring.

There are three primary SMS statuses: i) \emph{Sent}, which indicates that the mobile device has sent the SMS to the SMSC and the SMSC has confirmed its reception, ii) the \emph{Delivered}, meaning that the recipient has received the SMS and has responded with the Delivery Report, and iii) \emph{Failed} when errors occur.

\subsection{Network Delay Factors}\label{delays}

\begin{figure}[!t]
    \centering
    \includegraphics[width=1.\columnwidth,keepaspectratio]{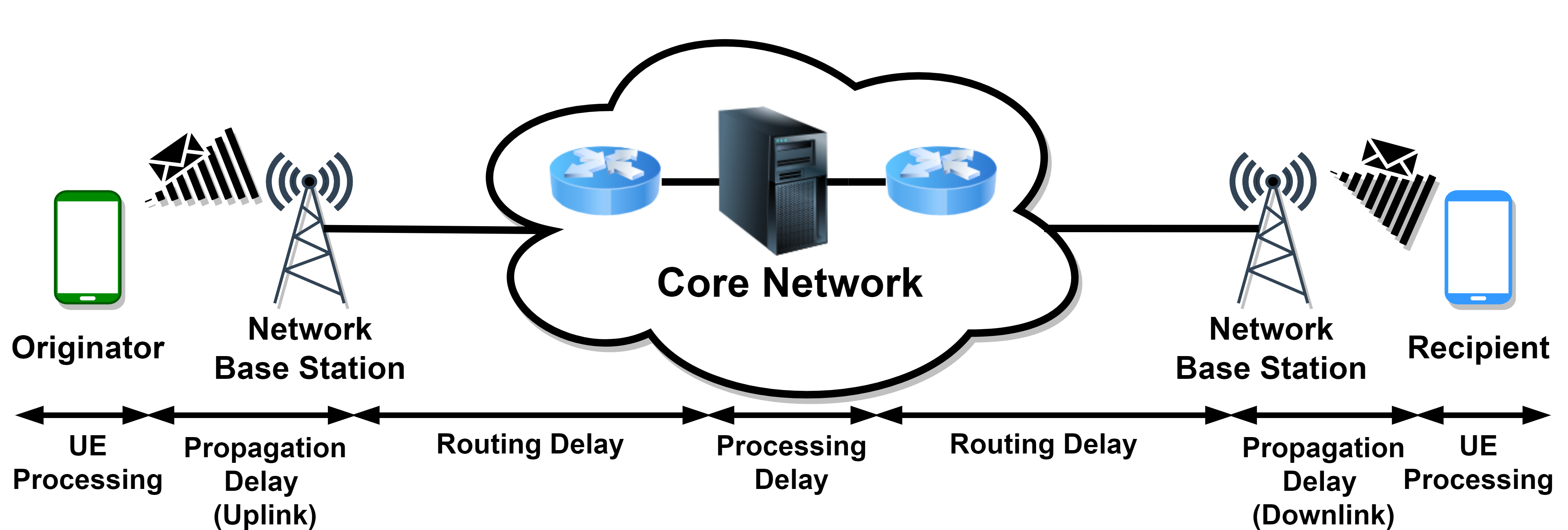}
    \caption{Various delays for one SMS transmission between two users. Similar delays apply for the Delivery Report which is sent back to the originator.}
    \label{fig:sys-delays}
\end{figure}

SMS text transmissions and Delivery Reports incur timing delays in the communication channel. Figure~\ref{fig:sys-delays} illustrates the delays for a single SMS transmission between the originator and the recipient.

\noindent \textbf{(1) UE Processing}: This is the time taken by the phone to process the SMS for transmission or reception. The corresponding base station has already completed its transmission at that time. The processing includes the modem and OS procedures, and the user-related services used (\eg, calls, SMSes, mobile data) that occupy uplink and downlink resources.

\noindent \textbf{(2) Propagation Delay}: This depends on the RAN network's design, configuration, and deployment including the front-haul of the mobile network, physical properties and quality of the signal, transmission capabilities of the base station, and management of the uplink and downlink communications.

\noindent \textbf{(3) Routing Delay}: SMS messages pass through multiple network entities depending on the architecture and the generation (\eg, LTE, 5G, \etc). 
Routing delays occur in the mobile back-haul, \ie the transport network that connects the core network and the RAN, as well as within the core network. 
Apart from the SMSC and the gateways, the SMS may require additional processing, \eg, by the AMF (5G SA), MME (LTE-5G NSA), and IMS, before reaching the destination, thereby also contributing to the routing delays.

\noindent \textbf{(4) Processing Delay}: This delay generally includes the SMSC, the IMS, and MSC/MME/AMF processing. The SMSC manages the SMS reception and delivery process and may also deploy congestion, filtering, and prioritization techniques.

\section{SMS-based Location Inference Attack}
\label{sec:attack-concept}

The high-level idea of our attack is as follows: The time elapsed between sending an SMS and receiving the corresponding SMS Delivery Report differs depending on the receiver's current location, implying one can distinguish different receiver locations by observing the elapsed time.

\subsection{Attacker Goal and Assumptions}

The attacker's goal is to locate the victim receiver's whereabouts, specifically, whether the victim's mobile is in a specific geographic area of interest.\footnote{We do not tackle the tracking of exact movement patterns of the victim in this paper.} 

We assume that the attacker knows the victim's mobile number and can send an SMS to that number. 
The SMS can be regular private messages, undirected mass messages (e.g., marketing, advertisements) that the victim will likely ignore, or a silent SMS that victim's device acknowledges without any content or alerts, remaining entirely unnoticed by the victim. 
We assume the attacker can target any subscriber (victim) with a valid mobile number attached to a cellular provider and maintain a typical connection to send text messages to the victim and receive delivery notifications. The adversary can access any network operator using the corresponding (e)SIM as a normal user.

Additionally, we assume the attacker can collect measurements from locations of interest directly from the victim when located at specific locations/areas of interest (without revealing the attack) or deploy similar devices and connections as the victim at these locations for data collection. The attacker is \emph{not} limited in terms of the number of smartphone devices, (e)SIMs, mobile numbers, or subscription plans. The attack does \emph{not} require physical access to the victim's USIM cards, mobile devices, or any network entities (\eg, base stations, core network, \etc). Finally, the attacker neither obtains nor modifies sensitive information, \eg, cryptographic keys.

\subsection{Timing Features} \label{sec:features}

\begin{figure}[tb]
    \centering
    \includegraphics[width=1.\columnwidth,keepaspectratio]{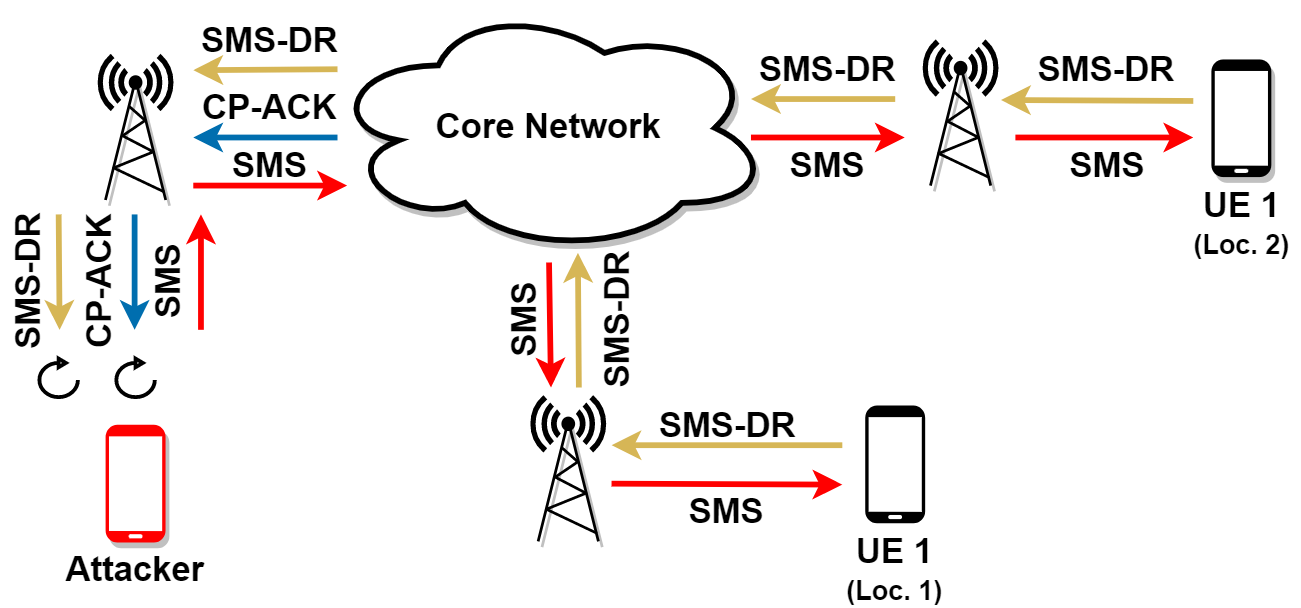}
    \caption{Network flow for SMS transmissions in different locations.}
    \label{fig:sms-burst}
\end{figure}

As shown in Figure~\ref{fig:sms-burst}, SMS transmissions to different device locations generate acknowledgments from the core network (CP-ACK) associated with \emph{Sent} notification and Delivery Reports (SMS-DR) from the receivers resulting in the \emph{Delivered} status. Hence, the attacker can leverage three timestamps to execute the attack: 

\begin{compactitem}
 \item $t_{tx}$: \texttt{SMS transmit time} as the time when attacker sends the SMS, 
 \item $t_{sent}$: \texttt{SMS sent time} as the time when the attacker receives the "Sent" notification, and 
 \item $t_{del}$: \texttt{SMS delivery time} as the time when the attacker receives the "Delivered" notification.
\end{compactitem}

After the SMS transmission is complete, the real sent duration $T_{sent}$, the real delivery duration $T_{del}$, the total delivery duration $T_{tot}$, and the delivery ratio $P$ can be calculated as follows:

\begin{equation} \label{feat_sent}
    T_{sent} = t_{sent} - t_{tx}
\end{equation}
\begin{equation} \label{feat_del}
    T_{del} = t_{del} - t_{sent}
\end{equation}
\begin{equation} \label{feat_tot}
    T_{tot} = T_{del} + T_{sent} 
\end{equation}
\begin{equation} \label{feat_ratio}
   P = \frac{T_{del}}{T_{tot}} = \frac{t_{del}-t_{sent}}{t_{del} - t_{tx}}
\end{equation}

\begin{figure}[tb]
    \centering
    \includegraphics[width=0.8\columnwidth,keepaspectratio]{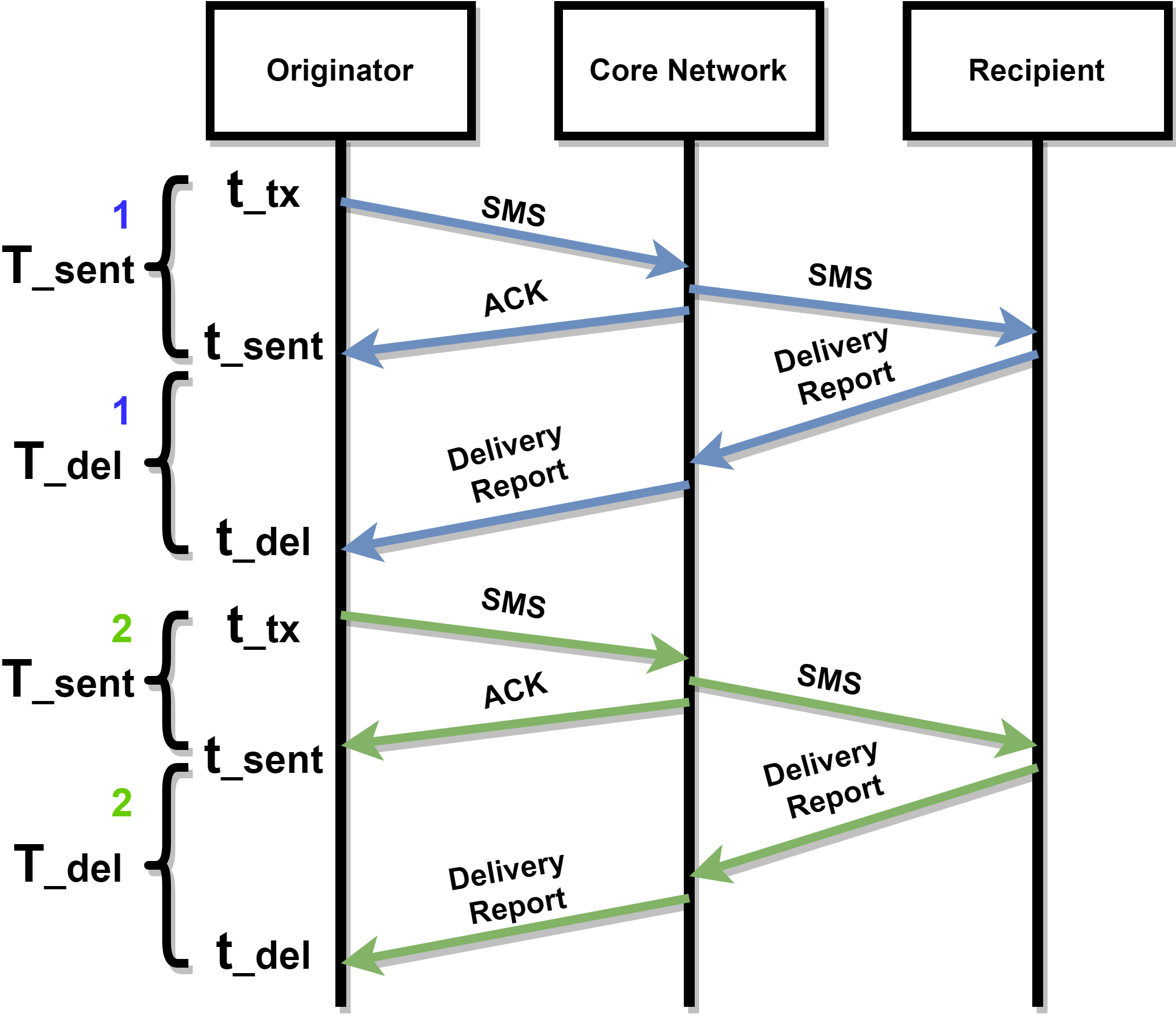}
    \caption{Timing features for each SMS transmission.}
    \label{fig:timings}
\end{figure}

These features apply to each individual SMS transmission only. Figure~\ref{fig:timings} shows the timing features for two SMS transmissions.

To produce robust location signatures and generate a pattern, we consider two consecutive SMS transmissions ($i-1$ and $i$), and estimate the difference in real sent duration $T_{\Delta sent}$ and real delivery duration $T_{\Delta del}$, respectively:\begin{equation}\label{feat_sent_diff}
    T_{\Delta sent} = (T_{sent}^i - T_{sent}^{i-1}) / T_{sent}^{i-1}
\end{equation}
\begin{equation}\label{feat_del_diff}
    T_{\Delta del} = (T_{del}^i - T_{del}^{i-1}) / T_{del}^{i-1}
\end{equation}

The \emph{location signature} is a combination of these six features: $(T_{sent}, T_{del}, T_{tot}, P, T_{\Delta sent}, T_{\Delta del})$. 

\subsection{Attack Concept}

The attack is conducted in two phases: (i) A \emph{Preparation} and (ii) an \emph{Attack} phase.

In the \emph{Preparation} phase, the adversary repeatedly sends multiple (silent) SMS, with Delivery Reports enabled, to the victim while observing their respective locations. 
The attacker collects measurements to identify the timing characteristics of the victim's locations.
Despite being aware of the victim's locations at this stage, the victim will not notice that they are being surveilled when the adversary uses silent SMSes. 
Using these measurements and analyzing the different timing features outlined in Section~\ref{sec:features}, fingerprints for each of the victim's locations are generated.

In the \emph{Attack} phase, the adversary collects new measurements without knowing the victim's location and attempts to determine their current location based on the timings. 
To do this, the adversary must solve a \emph{classification} problem, \ie, assign the newly observed measurements to one of the previously seen locations by comparing timings with the respective location fingerprints.
Depending on the victim's movement patterns and the locations observed in the preparation phase, the classification occurs in multiple iterations.
Therefore, the classification problem is partitioned into a step-wise location prediction problem involving several location identification tasks with decreasing granularity levels from classifying international locations to regional (\eg, at city-level).

\begin{figure}[tb]
    \centering
    \includegraphics[width=1.\columnwidth,keepaspectratio]{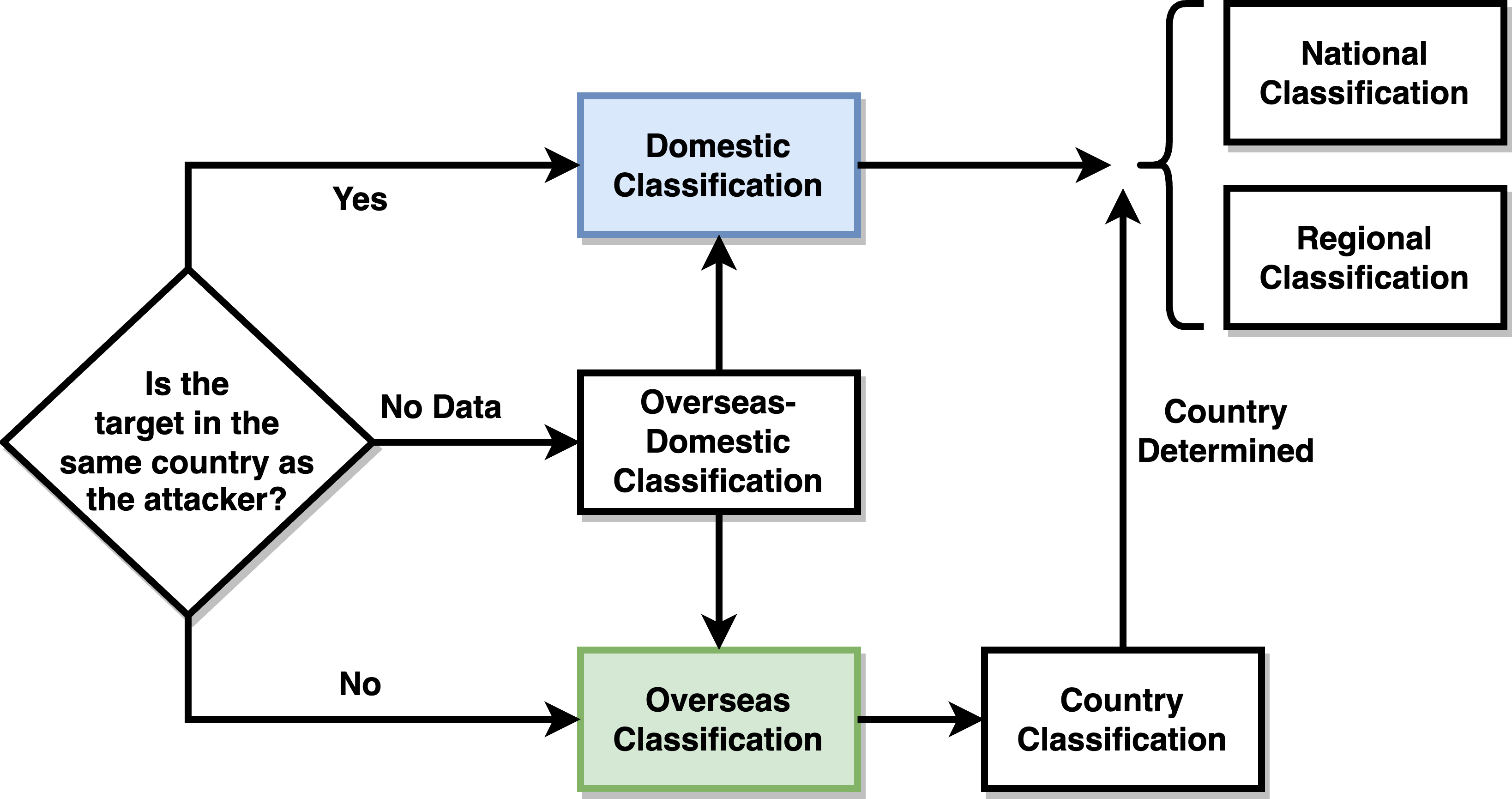}
    \caption{The classification methodology}
    \label{fig:class-method}
\end{figure}

\noindent\textbf{Classification Methodology.} We describe the classification approach that the attacker follows to retrieve a victim's location in multiple iterations (Figure~\ref{fig:class-method}). We use the example of a victim moving internationally.

Initially, the attacker may not have sufficient intelligence regarding the victim's current country of residence.
Thus, the first step is to determine whether the victim is \emph{Overseas} or \emph{Domestic}. If the victim is overseas, then the attacker proceeds with determining the specific country (country-based classification). 
Once the country is known, the attacker may choose to perform either a national or regional classification depending on the attacker's objectives and the victim's routine. In the regional classification, the attacker attempts to discover the victim's location within a limited area, while the national classification has a macroscopic view of the country, incorporating cities and towns.

Having knowledge about the victim's general geographical whereabouts such as North America, can help narrow down potential candidate locations making classification more manageable. If there is only one country and one city, the methodology can be simplified to just regional location identification. Therefore, the attacker does not need to adhere to the entire methodology as it primarily depends on the victim's routine.

\begin{figure*}[tb]
    \centering
    \includegraphics[width=0.9\textwidth]{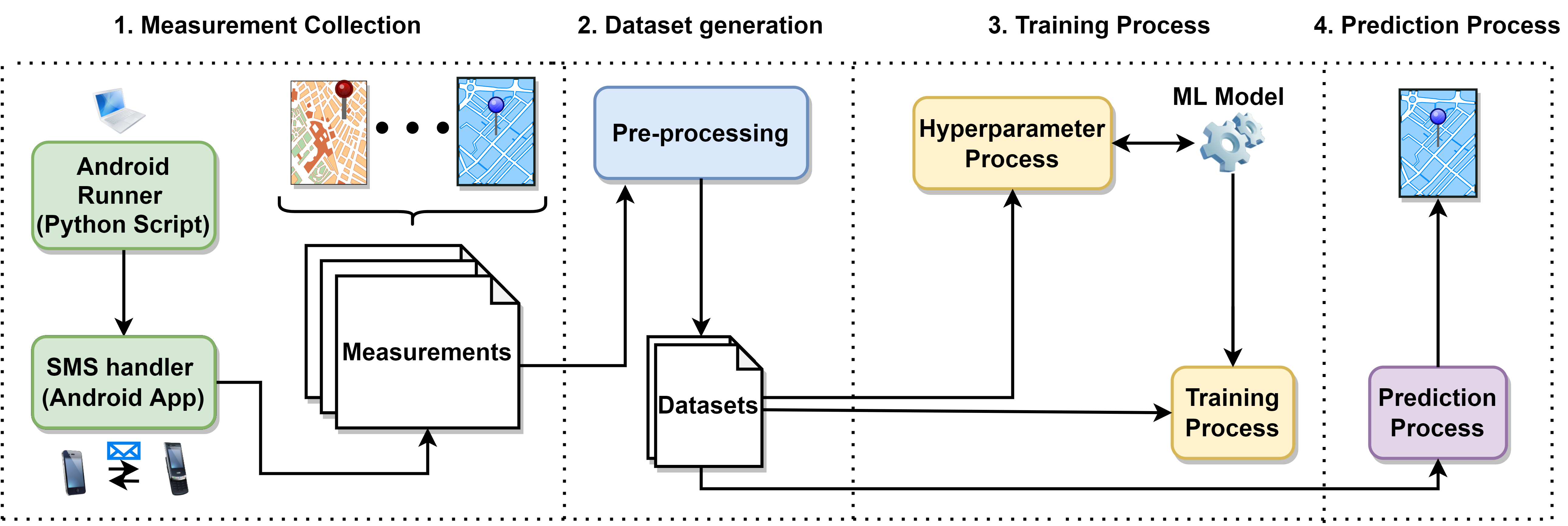}
    \caption{Components and stages of the SMS location identification attack.}
    \label{fig:sys-comp}
\end{figure*}

\section{Experimental Validation}
\label{sec:experimental-validation}

In this section, we present our experimental validation of the SMS-based location inference attack.
We describe our measurement setup and the different stages of the two attack phases outlined in Section~\ref{sec:attack-concept}.
Figure~\ref{fig:sys-comp} provides an overview of our experimental procedure.

\subsection{Setup}\label{setup}

We send SMSes between smartphones at different geographical locations to collect measurements for our experiments.
Our setup includes \textit{active devices} (phones) controlled via the Android Debug Bridge (ADB) to send SMSes to other devices.
These phones are configured to analyze cellular traffic and baseband logs to extract timing and network information such as protocols, connections with the core network, AT SIM commands, \etc 
Active devices have SMS Delivery Reports enabled to visualize notifications while sending messages. \textit{Passive devices} are used to receive messages. We list all the devices used in Table~\ref{tab:specs}.

Our devices are located across several countries, including the United States (US), UAE (AE), and seven countries in Europe (BE, DE, DK, GR, LU, NL, UK). The experiments cover ten operators and several generation technologies such as LTE, LTE+, 5G NSA/SA. Additionally, we record the approximate channel condition such as strength and quality, for each receiving location. Table~\ref{tab:location-characteristics} presents the relevant characteristics of all locations that appear in our measurements.

We conduct three rounds of measurements serving different purposes:

\begin{compactenum}[(i)]
\item We conduct long-distance international measurements with devices in multiple countries, with the sender located in AE-1. 

\item We send messages from a single active device to passive devices at various domestic locations, including multiple cities and locations within them, for AE, GR, DE, NL, BE, and LU. The experiments are conducted from different sender locations, with the sender in AE-1 for AE experiments, GR-1 for the GR experiments, and DE-4 for the rest. The primary objective is to demonstrate a practical and realistic scenario involving a person's natural everyday behavior on a smaller scale including regular commuting to adjacent countries.

\item We collected measurements across different operators and roaming devices at several locations. Specifically, we focused on distinguishing between network operators and smartphone devices which assists our location identification.
\end{compactenum}

\begin{table}[tb!]
\centering
\caption{\label{tab:location-characteristics} Receiver locations and their characteristics. GR-1, AE-1 and DE-4 acted as senders (using LTE/LTE+/5G) and receivers, but the table focuses on the receivers only. Channel conditions show the approximate connection quality from our devices in those locations. Receivers that have ranges in \textit{Distance} column represent an area instead of a specific fixed position.}
\resizebox{\columnwidth}{!}{%
\begin{tabular}{@{}l@{}rl@{\hspace{0.1cm}}ccl@{}}
\toprule
\textbf{Rec.} & Dist. [km] & Connection Type & Routing & Cond. & Operator\\
\midrule
\multicolumn{6}{l}{\emph{International Receiver Locations}}\\
\textbf{Int-GR} & 3266  & LTE,LTE+ & SMSoIP & \includegraphics[width=0.03\textwidth, height=4mm]{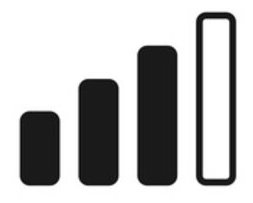} & C\\
\textbf{Int-DE} & 5460 & LTE,LTE+ & SMSoIP & \includegraphics[width=0.03\textwidth, height=4mm]{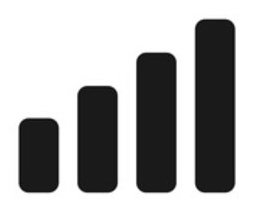} & E\\
\textbf{Int-DK} & 5880 & LTE+,5G NSA/SA & SMSoIP & \includegraphics[width=0.03\textwidth, height=4mm]{images/bars/4-bars.jpeg} & I\\
\textbf{Int-UK} & 5700 & 5G NSA/SA & SMSoIP & \includegraphics[width=0.03\textwidth, height=4mm]{images/bars/4-bars.jpeg} & H\\
\textbf{Int-US} & 10.710 & LTE,LTE+ & SMSoIP & \includegraphics[width=0.03\textwidth, height=4mm]{images/bars/3-bars.jpeg} & J\\
\midrule
\multicolumn{6}{l}{\emph{Receiver Locations in the UAE}}\\
\textbf{AE-1} & \shortstack{1-7 meters} & 5G NSA/SA & SMSoIP & \includegraphics[width=0.03\textwidth, height=4mm]{images/bars/4-bars.jpeg} & A, C\\
\textbf{AE-2} & \shortstack{10} & 5G NSA/SA & SMSoIP & \includegraphics[width=0.03\textwidth, height=4mm]{images/bars/3-bars.jpeg} & A, B\\ 
\textbf{AE-3} & \shortstack{14} & LTE,LTE+ & SMSoIP & \includegraphics[width=0.03\textwidth, height=4mm]{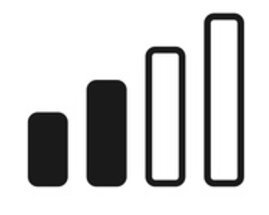} & A, B\\ 
\textbf{AE-4} & \shortstack{135} & LTE+,5G SA & SMSoIP & \includegraphics[width=0.03\textwidth, height=4mm]{images/bars/4-bars.jpeg} & A\\
\midrule
\multicolumn{6}{l}{\emph{Receiver Locations in Greece}}\\
\textbf{GR-1} & \shortstack{1-5 meters} & LTE,LTE+ & SMSoIP & \includegraphics[width=0.03\textwidth, height=4mm]{images/bars/3-bars.jpeg}  & C, D\\
\textbf{GR-2} & \shortstack{8} & LTE,LTE+ & SMSoIP & \includegraphics[width=0.03\textwidth, height=4mm]{images/bars/3-bars.jpeg} & C\\ 
\textbf{GR-3} & \shortstack{12} & LTE,LTE+ & SMSoIP & \includegraphics[width=0.03\textwidth, height=4mm]{images/bars/4-bars.jpeg} & C\\
\textbf{GR-4} & \shortstack{180} & LTE & SMSoIP & \includegraphics[width=0.03\textwidth, height=4mm]{images/bars/3-bars.jpeg} & C\\
\textbf{GR-5} & \shortstack{200} & LTE & SMSoIP & \includegraphics[width=0.03\textwidth, height=4mm]{images/bars/2-bars.jpeg} & C\\
\textbf{GR-6} & \shortstack{290} & LTE & SMSoIP & \includegraphics[width=0.03\textwidth, height=4mm]{images/bars/3-bars.jpeg} & C\\
\midrule
\multicolumn{6}{l}{\emph{Receiver Locations in Germany}}\\
\textbf{DE-1} & \shortstack{11} & LTE,LTE+ & SGsAP/Diameter & \includegraphics[width=0.03\textwidth, height=4mm]{images/bars/4-bars.jpeg} & E,F,G \\
\textbf{DE-2} & \shortstack{45} & LTE,LTE+ & SGsAP/Diameter & \includegraphics[width=0.03\textwidth, height=4mm]{images/bars/3-bars.jpeg}  & E,F,G \\
\textbf{DE-3} & \shortstack{2} & LTE,LTE+ & SGsAP/Diameter & \includegraphics[width=0.03\textwidth, height=4mm]{images/bars/4-bars.jpeg} & E,F,G \\
\textbf{DE-4} & \shortstack{0} & LTE,LTE+ & SGsAP/Diameter & \includegraphics[width=0.03\textwidth, height=4mm]{images/bars/3-bars.jpeg}  & E,F,G \\
\textbf{DE-5} & \shortstack{31} & LTE,LTE+ & SGsAP/Diameter & \includegraphics[width=0.03\textwidth, height=4mm]{images/bars/4-bars.jpeg} & E,F,G \\
\textbf{DE-6} & \shortstack{0~--~5} & LTE,LTE+ & SGsAP/Diameter & \includegraphics[width=0.03\textwidth, height=4mm]{images/bars/4-bars.jpeg} & E,F,G \\
\textbf{DE-7} & \shortstack{0~--~35} & LTE,LTE+ & SGsAP/Diameter & \includegraphics[width=0.03\textwidth, height=4mm]{images/bars/3-bars.jpeg} -- \includegraphics[width=0.03\textwidth, height=4mm]{images/bars/4-bars.jpeg} & E,F,G\\
\textbf{DE-8} & \shortstack{110~--~130} & LTE,LTE+ & SGsAP/Diameter & \includegraphics[width=0.03\textwidth, height=4mm]{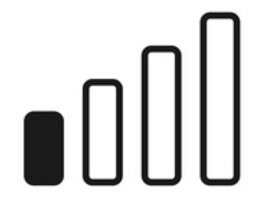} -- \includegraphics[width=0.03\textwidth, height=4mm]{images/bars/4-bars.jpeg} & E,F,G\\
\textbf{DE-9} & \shortstack{0~--~110} & LTE,LTE+ & SGsAP/Diameter & \includegraphics[width=0.03\textwidth, height=4mm]{images/bars/1-bar.jpeg} -- \includegraphics[width=0.03\textwidth, height=4mm]{images/bars/4-bars.jpeg} & E,F,G\\
\textbf{DE-10} & \shortstack{59} & LTE,LTE+ & SGsAP/Diameter & \includegraphics[width=0.03\textwidth, height=4mm]{images/bars/2-bars.jpeg} & E,F,G\\
\midrule
\multicolumn{6}{l}{\emph{Receiver Locations in the Netherlands}}\\
\textbf{NL-1} & \shortstack{130} & LTE,LTE+ & SGsAP/Diameter & \includegraphics[width=0.03\textwidth, height=4mm]{images/bars/3-bars.jpeg} & E,G\\
\textbf{NL-2} & \shortstack{125} & LTE,LTE+ & SGsAP/Diameter & \includegraphics[width=0.03\textwidth, height=4mm]{images/bars/4-bars.jpeg} & G\\
\textbf{NL-3} & \shortstack{90} & LTE,LTE+ & SGsAP/Diameter & \includegraphics[width=0.03\textwidth, height=4mm]{images/bars/4-bars.jpeg} & G \\
\textbf{NL-4} & \shortstack{129} & LTE,LTE+ & SGsAP/Diameter & \includegraphics[width=0.03\textwidth, height=4mm]{images/bars/4-bars.jpeg} & E,F,G \\
\textbf{NL-5} & \shortstack{128~--~130} & LTE,LTE+ & SGsAP/Diameter & \includegraphics[width=0.03\textwidth, height=4mm]{images/bars/4-bars.jpeg} & E,F,G \\
\midrule
\multicolumn{6}{l}{\emph{Receiver Locations in Belgium}}\\
\textbf{BE-1} & \shortstack{195} & LTE,LTE+ & SGsAP/Diameter & \includegraphics[width=0.03\textwidth, height=4mm]{images/bars/3-bars.jpeg} & E,F,G\\
\textbf{BE-2} & \shortstack{153} & LTE,LTE+ & SGsAP/Diameter & \includegraphics[width=0.03\textwidth, height=4mm]{images/bars/3-bars.jpeg} & E,F,G\\
\textbf{BE-3} & \shortstack{116~--~210} & LTE,LTE+ & SGsAP/Diameter & \includegraphics[width=0.03\textwidth, height=4mm]{images/bars/1-bar.jpeg} -- \includegraphics[width=0.03\textwidth, height=4mm]{images/bars/4-bars.jpeg} & E,F,G\\
\midrule
\multicolumn{6}{l}{\emph{Receiver Locations in Luxembourg}}\\
\textbf{LU-1} & \shortstack{220} & LTE,LTE+ & SGsAP/Diameter & \includegraphics[width=0.03\textwidth, height=4mm]{images/bars/4-bars.jpeg} & E,F,G\\
\textbf{LU-3} & \shortstack{165~--~225} & LTE,LTE+ & SGsAP/Diameter & \includegraphics[width=0.03\textwidth, height=4mm]{images/bars/1-bar.jpeg} -- \includegraphics[width=0.03\textwidth, height=4mm]{images/bars/4-bars.jpeg} & E,F,G\\
\bottomrule
\multicolumn{6}{l}{\textbf{Locations (Cities/Regions):} \emph{Int-GR}: Athens, \emph{Int-DE}: Bochum, \emph{Int-DK}: Copenhagen,}\\
\multicolumn{6}{l}{\emph{Int-UK}: London, \emph{Int-US}: Boston, \emph{AE-1,3}: Abu Dhabi, \emph{AE-2}: Saadiyat, \emph{AE-4}: Dubai,}\\
\multicolumn{6}{l}{\emph{GR-1,2,3}: Athens, \emph{GR-4}: Chania, \emph{GR-5}: Messenia, \emph{GR-6}: Thessaloniki,}\\
\multicolumn{6}{l}{\emph{DE-1}: Dortmund, \emph{DE-2}: Raesfeld, \emph{DE-5}: Unna, \emph{DE-3,4,6}: Bochum, }\\
\multicolumn{6}{l}{\emph{DE-7}: Ruhr Area, \emph{DE-8}: Aachen Area, \emph{DE-9}: NRW State, \emph{DE-10}: Borken,}\\
\multicolumn{6}{l}{\emph{NL-1,4,5}: Veldhoven, \emph{NL-2}: Eindhoven, \emph{NL-3}: Roermond}\\
\multicolumn{6}{l}{\emph{BE-1}: Bastogne, \emph{BE-2}: Sankt-Vith, \emph{BE-3}: Wallonia Region}\\
\multicolumn{6}{l}{\emph{LU-1}: Luxembourg City, \emph{LU-3}: Western Regions}\\
\multicolumn{6}{l}{\textbf{Operators:} \emph{A}: Etisalat (UAE), \emph{B}: du (UAE), \emph{C}: Cosmote (GR),}\\
\multicolumn{6}{l}{\emph{D}: Vodafone (GR) \emph{E}: Telekom (DE), \emph{F}: Vodafone (DE), \emph{G}: Telefonica (DE),}\\
\multicolumn{6}{l}{\emph{H}: Vodafone (UK), \emph{I}: Telenor (DK), \emph{J}: Mint (US)}
\end{tabular}
}
\end{table}

\subsection{Measurement Collection}\label{measurement-collection}

Our data collection is sketched in step~1 of Figure~\ref{fig:sys-comp}.
We developed an Android application called \textit{SMS handler} that runs on active devices and sends one silent SMS at a time to a target device.
Once the SMS is sent, the application waits for the Delivery Report (both \emph{Sent} and \emph{Delivered} notifications) and records all the required timestamps and computes the features~(\ref{feat_sent})-(\ref{feat_ratio}) (\cf~Section~\ref{sec:features}).

We use a python script, \textit{Android Runner}, to automate SMS transmission to a designated receiver and capture the Delivery Report timings for each SMS.
The script interacts with the smartphone through basic ADB commands and key events (to press buttons, fill text input fields, etc.) without requiring device rooting.
The script runs on a Dell Latitude E5450 and a regular desktop computer (\cf~step 1 in Figure~\ref{fig:sys-comp}) using a \emph{cronjob} for repeated execution.

We schedule \textit{SMS burst}, \ie, consecutive 20 SMS transmissions, on an hourly basis. 
To distribute the SMSes for each location, we span them over 2 to 3 days to avoid potential SMS spam filtering and prevent network congestion, which may affect the timings. 
This procedure also helps us collect representative traffic dataset, including various times of the day, potential network configuration changes, and different levels of network loads. Throughout our measurement campaign, we have sent and accumulated around 155,512 SMSes. 
Refer to Table~\ref{tab:counts} for SMS numbers per device, per country, and per operator.

We constantly monitor whether the active device sent the silent SMS successfully during our experiments. 
We use the Android logging tool \textit{Logcat} to investigate the routing methods and connection establishments and track the SMS procedures. Appendix~\ref{baseband-log} provides further information.

\subsection{Dataset Generation}

We now describe how we aggregate our collected measurements to generate the evaluation dataset (step~2 in Figure~\ref{fig:sys-comp}).
We calculate the timing features from the collected data, generate location signatures, each composed of all six timing features obtained during or derived from a single measurement iteration (Section~\ref{sec:features}).

Our evaluation dataset contains signatures for each candidate location, covering various granularity levels, from domestics and overseas to national and regional classifications (\cf~Figure~\ref{fig:class-method}). In our data, we also identify the SMS routing modes, \ie, SMSoIP for LTE/LTE+, SMSoIP for 5G, and SGsAP/Diameter for LTE/LTE+.

\subsection{Location Classification}

We opted for \texttt{Multilayer Perceptron (MLP)} using Python's SKLearn libraries as classifier to perform location classification because of its flexibility in parametrization and high performance on large datasets. The model comprises a stochastic gradient descent solver, softmax and sigmoid activations for multiclass and binary classifications respectively, and three layers with 10, 40, and 10 nodes respectively for the input, hidden, and output layers.
Additionally, we set the maximum iterations to 5000, the learning rate to be constant, batch size to be 32, and the alpha to 0.0001. 
We performed automatic and manual parameter tuning to improve the model's accuracy (Appendix~\ref{tuning} provides more details). 
We focus on \emph{accuracy} throughout our classifications, measuring the number of correct predictions out of the total predictions made.

The training and prediction procedures correspond to steps~3 and~4 in Figure~\ref{fig:sys-comp}. The datasets are randomly split, while the class with the highest probability is assigned by the MLP classifier as the prediction result. Training and prediction processes utilize the cross-validation methodology with 10 k-folds to prevent over-fitting and promote model generalisation. We also compared the performance of a Random Forest Classifier, Decision Tree Classifier, and Recurrent Neural Network with Keras libraries, but the optimized MLP outperformed them all. Therefore, we present our results for the MLP model only.

\begin{table*}[tb!]
\centering
\caption{\label{tab:class-results-country} Classification results for international experiments.}
\resizebox{0.85\textwidth}{!}{%
\begin{tabular}{@{}lcllcc@{}}
\toprule
\textbf{Classification} & Size/Class & Operators & Receiver Locations & Sender Location & Accuracy \\
\midrule
\textbf{Overseas-vs.-Domestic} & 1200 & A, C, E, H, I, J & AE-X, Int-X & AE-1 & 96\% \\
\textbf{All Country-based} & 280 & C, E, H, I, J & Int-X & AE-1  & 96\% \\ 
\textbf{EU Country-based} & 280 & C, E, I & Int-GR, Int-DE, Int-DK & AE-1 & 95\% \\ 
\textbf{EU Country-based} & 257 & G & DE-4, NL-4, BE-1, LU-1 & DE-4 & 75\% \\
\textbf{EU Country-based} & 319 & E & DE-4, NL-4, BE-1, LU-1 & DE-4 & 74\% \\
\textbf{EU Country-based} & 313 & F & DE-4, NL-4, BE-1, LU-1 & DE-4 & 62\% \\
\bottomrule
\end{tabular}
}
\end{table*}

\section{Location Classification Results}\label{results}

We follow the classification methodology outlined in Section~\ref{sec:attack-concept} proceeding step-by-step from coarse- to fine-grained location classifications and present our results.

\subsection{International Classification}
\label{sec:classification-intl}

For the international classification, we focus on large geographical areas of the victim, primarily attempting to identify locations in different countries. Our results are shown in Table~\ref{tab:class-results-country}.

\begin{figure}
    \centering
    \includegraphics[width=\columnwidth]{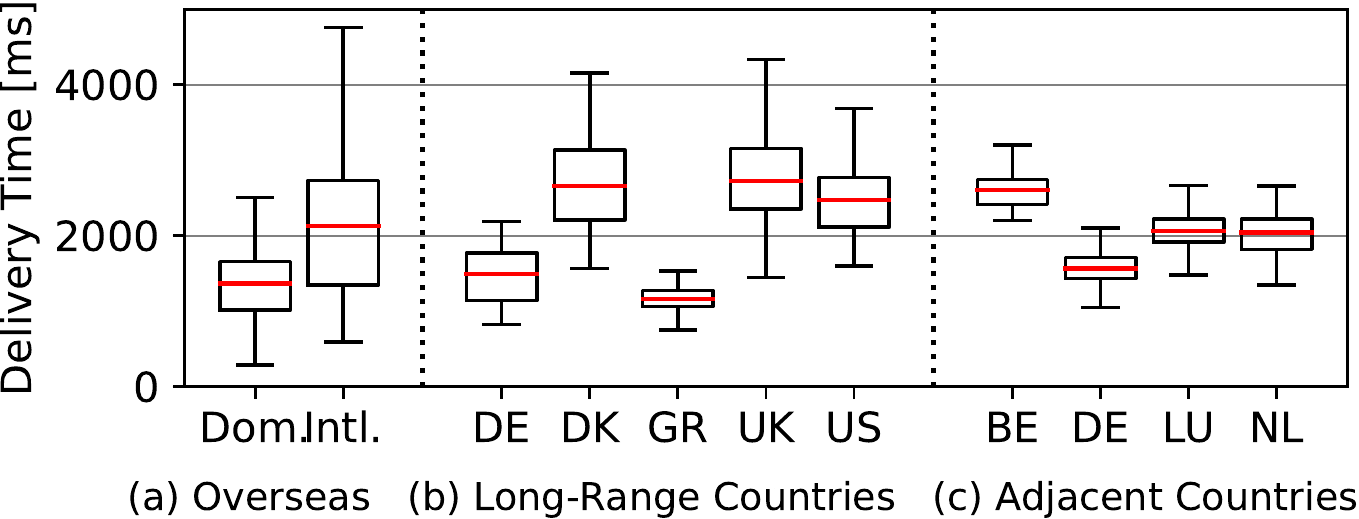}
    \caption{Delivery timings for receivers in various countries.}
    \label{fig:boxplots-countries}
\end{figure}

\textbf{Overseas-vs.-Domestic Classification} aims to determine whether the victim is within the home country or abroad. This binary classification experiment groups the AE locations (home country) together and Int-X locations together. The results indicate that the target can be identified with an accuracy of 96\%. The two box plots in Figure~\ref{fig:boxplots-countries}a show a clear timing difference between the two classes based on the Delivery Report ($T_{del}$), facilitating accurate identification.

\textbf{Country-based Classifications} aim to determine the victim's location in a specific country. First, we conduct experiments on countries that are far apart to demonstrate the existence of timing differences. We perform multi-class classification for all Int-X locations in different countries and achieve 96\% accuracy. The box plots in Figure~\ref{fig:boxplots-countries}b depict the timing difference in the dataset between GR, DE, DK, UK, and US locations. Next, we select only EU countries for a multi-class classification to identify locations within a smaller geographical area. We used Int-GR, Int-DE, Int-DK locations (sender AE-1, based on another continent) achieving 95\% accuracy. 

\begin{figure}[tb]
    \centering
    \includegraphics[width=\columnwidth, keepaspectratio]{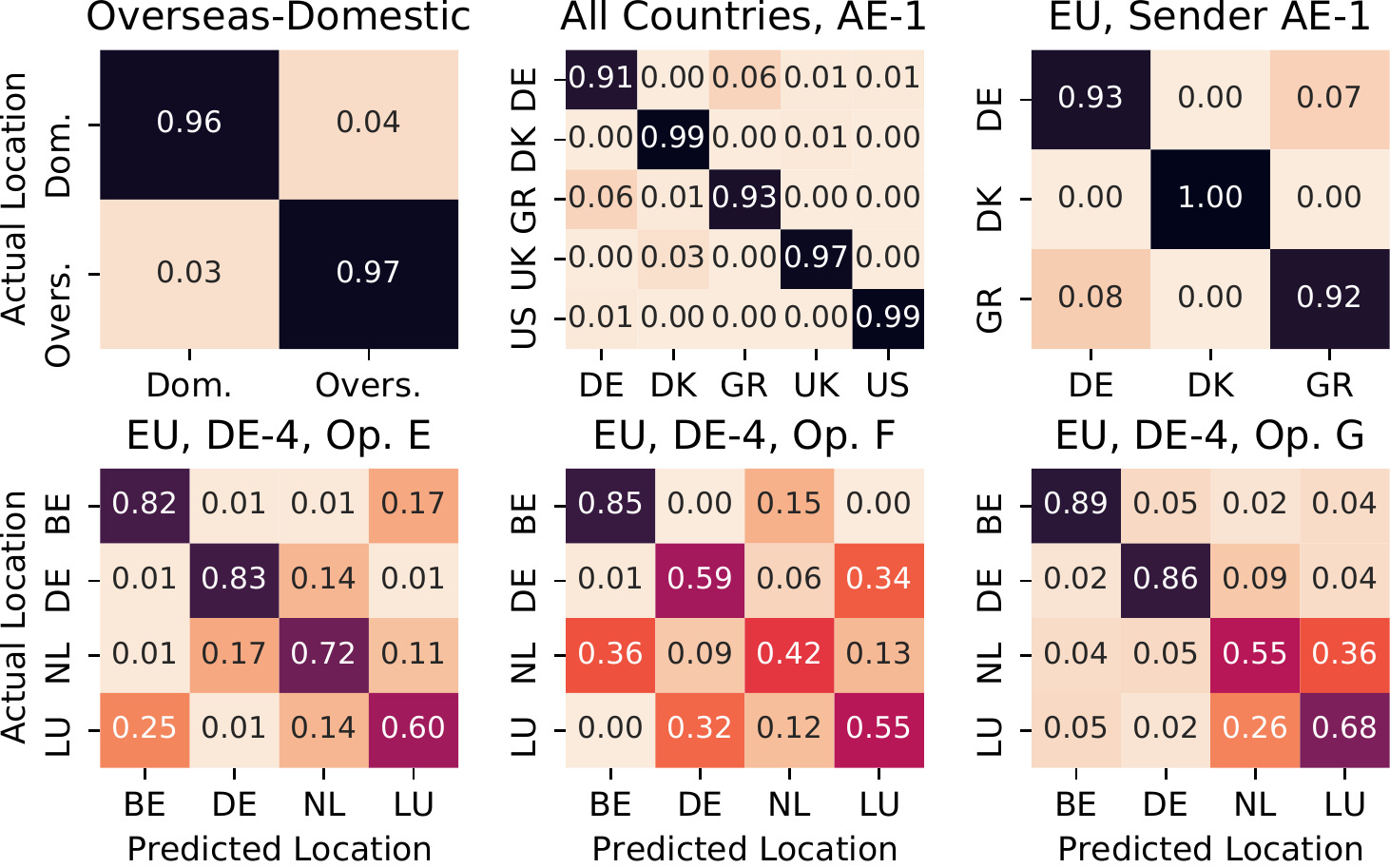}
    \caption{Confusion matrices for international classifications, displaying the results from Table~\ref{tab:class-results-country} in more detail. Although the model misclassifies more often for operator F, it achieves high accuracy in many cases.}
    \label{fig:int-confusion-matrices}
\end{figure}

In Figure~\ref{fig:int-confusion-matrices}, we present the confusion matrices for the overseas-vs.-domestic and country-based classifications (from Table~\ref{tab:class-results-country}). The figure confirms the high-accuracy results from the table and identifies the predictions that lead to less accurate results, involving classification with sender DE-4 and nearby receiver countries. For operators G and E, LU and NL receiver locations result in higher misclassifications than for DE and BE. The model also shows a loss of accuracy for operator F, where timing characteristics for DE, LU, and NL cause errors due to similarities, but the most likely returned result is still the correct one for each case. 

Finally, we performed a country-based classification targeting adjacent and nearby countries to identify even closer geographical locations. 
The victim traveled to DE-4, NL-4, BE-1, and LU-1 using operators G, E, and F. Our classifiers achieved 75\%, 74\%, and 62\% accuracy for these specific locations using operators G, E, and F, respectively. These three EU country-based classifications with four classes have an average accuracy of 70\% with the best performing being 75\% for operator G and E. Figure~\ref{fig:boxplots-countries}c shows the timing difference between those countries with NL-4 and LU-1 having similar delivery timings. However, raw delivery timing is only one of the six features we take into consideration in this case. 

\subsection{National \& Regional Classification}\label{national-regional}

In this section, we explore the location characteristics at a regional scale within the same country. Our classifications include receiver locations from Table~\ref{tab:location-characteristics}, which can be either fixed locations or areas through which a receiver is moving. We evaluate our model against fixed locations, areas, and their combination within each country. 
We repeated the classification for every combination of locations in our dataset, with sample sizes varying from 100 to 500. Table~\ref{tab:system-performance-national-regional} summarizes our results, broken down by receiver country and the number of locations, and includes the repetition with the largest sample size, which depends on the available data for each location.

\noindent\textbf{Fixed Locations.} Our classification achieves an average performance of 68\,\% in Germany based on 57 classifications of pairs of two locations. 
However, performance varies depending on the pairs of locations, so the average must be interpreted carefully.
The best performing classification (DE-3 and DE-5) achieves 92\,\% classification accuracy. 
Detailed results for all pairs of locations in Germany are presented by the matrix in Table~\ref{tab:any-matrix-de}. 
The average performance for the Netherlands across 15 classifications of location pairs is 63\,\%, with 98\,\% classification accuracy for NL-2 and NL-3. For Belgium, the overall performance is 86\,\%, but this only includes four classifications of the same two locations (BE-1 and BE-2) 40\,km apart from each other, using different phones.

\begin{table}[tb!]
\centering
\caption{\label{tab:system-performance-national-regional} Summary of regional/national classifications within the same country.}
\resizebox{.99\columnwidth}{!}{%
\begin{tabular}{@{}lcccc@{}}
\toprule
\textbf{Type} & \multicolumn{2}{c}{\textbf{All Classifications}} & \multicolumn{2}{c}{\textbf{Best Performing}}\\
& Num$^\ast$ & Avg. Acc. & Loc. Set & Accuracy \\
\midrule
\multicolumn{5}{l}{\emph{Regional classifications with 2 locations} (Random: 50\,\%)}\\
DE Fixed & 57 & 68\,\%   & DE-\{3,5\} & 92\,\% \\ 
NL Fixed & 15 & 63\,\% & NL-\{2,3\} & 98\,\% \\ 
BE Fixed & 4 & 86\,\% & BE-\{1,2\} & 95\,\% \\ 
\midrule
DE Area & 21 & 57\,\% & DE-\{6,8\} & 72\,\%  \\ 
\midrule
DE Mixed & 80 & 67\,\% & DE-\{8,10\} & 88\,\% \\ 
NL Mixed & 4 & 71\,\% & NL-\{3,5\} & 88\,\% \\ 
 BE Mixed & 8 & 77\,\% & BE-\{2,3\} & 84\,\% \\ 
 LU Mixed & 4 & 67\,\% & LU-\{1,3\} & 72\,\% \\ 
\midrule
\midrule
\multicolumn{5}{l}{\emph{Regional classification with 3 locations} (Random: 33\,\%)}\\
AE Fixed & 1 & 76\,\% & AE-\{1,2,3\} & 76\,\% \\
GR Fixed & 2 & 79\,\% & GR-\{1,2,3\} & 82\,\% \\
DE Fixed & 46 & 54\,\% & DE-\{2,5,10\} & 83\,\% \\ 
NL Fixed & 7 & 48\,\% & NL-\{1,2,3\} & 68\,\% \\ 
\midrule
DE Area & 13 & 41\,\% & DE-\{6,7,8\} & 50\,\% \\ 
\midrule
DE Mixed & 252 & 50\,\% & DE-\{5,6,10\} & 81\,\% \\ 
NL Mixed & 6 & 59\,\% & NL-\{1,3,5\} & 78\,\% \\ 
BE Mixed & 4 & 67\,\% & BE-\{1,2,3\} & 73\,\% \\ 
\midrule
\midrule
\multicolumn{5}{l}{\emph{Regional classification with 4 locations} (Random: 25\,\%)}\\
AE Fixed & 1 & 58\,\% & AE-\{1,2,3,4\} & 58\,\% \\
DE Fixed & 19 & 47\,\% & DE-\{1,2,5,10\} & 64\,\% \\ 
NL Fixed & 1 & 53\,\% & NL-\{1,2,3,4\} & 53\,\% \\ 
\midrule
DE Area & 3 & 34\,\% & DE-\{6,7,8,9\} & 38\,\% \\ 
\midrule
DE Mixed & 402 & 41\,\% & DE-\{2,5,9,10\} & 67\,\% \\ 
NL Mixed & 4 & 48\,\% & NL-\{1,2,3,5\} & 58\,\% \\ 
\midrule
\midrule
\multicolumn{5}{l}{\emph{Regional classification with 5 locations} (Random: 20\,\%)}\\
DE Fixed & 3 & 37\,\% & DE-\{2,3,4,5,10\} & 50\,\% \\ 
\midrule
DE Mixed & 398 & 34\,\% & DE-\{2,3,5,8,10\} & 55\,\% \\ 
NL Mixed & 1 & 42\,\% & NL-\{1,2,3,4,5\} & 42\,\% \\ 
\bottomrule
\bottomrule
\end{tabular}
}
\raggedright\small
$^\ast$\emph{Num} denotes the numbers of different classifications, \eg, for different sets of receiver locations and phones.
\end{table}

Our classification scores decrease for larger sets of locations in all countries, but it should be noted that the chance of randomly guessing the correct location is also lower (\eg, 33\,\% for 3 locations instead of 50\,\% for 2 locations). Nevertheless, the average classification scores of 76\,\% and 79\,\% in the UAE and in Greece, respectively, still indicate a high performance. 

\noindent\textbf{Areas with Multiple Locations.} Areas can be challenging to distinguish as they are not associated with the attributes of one location only and may overlap. We report area classification results for DE locations in Table~\ref{tab:any-matrix-de}. In binary classifications, the model achieves an average accuracy of 57\,\% for 21 classifications, with DE-6 and DE-8 being the best-performing pair reaching 72\,\%. For three and four classes in DE, the model achieves 41\,\% and 34\,\%, respectively. Similar to the fixed locations, performances should be read and understood separately, as each combination has different features.

\noindent\textbf{Mixed Locations.} In this scenario, we explore the combinations of fixed locations and areas, which shows that the attacker is not limited to distinct types only. We used measurements from DE, NL, BE, and LU for the classification tasks in Table~\ref{tab:system-performance-national-regional}. In binary classifications, the model achieves 67\,\%, 71\,\%, 77\,\% and 67\,\% on average for DE, NL, BE and LU locations, respectively, while reaching up to 88\,\% in certain classifications. The model scores lower for classifications that include three, four, and five locations. For example, DE has an average accuracy of 50\,\%, 41\,\%, and 34\,\% for three, four, and five classes, respectively. Nonetheless, the large number of classifications with even diverse features should be taken into account cautiously, \ie, 252, 402, and 398 for three, four, and five classes, respectively.

\begin{figure}
    \centering
    \includegraphics[width=\columnwidth]{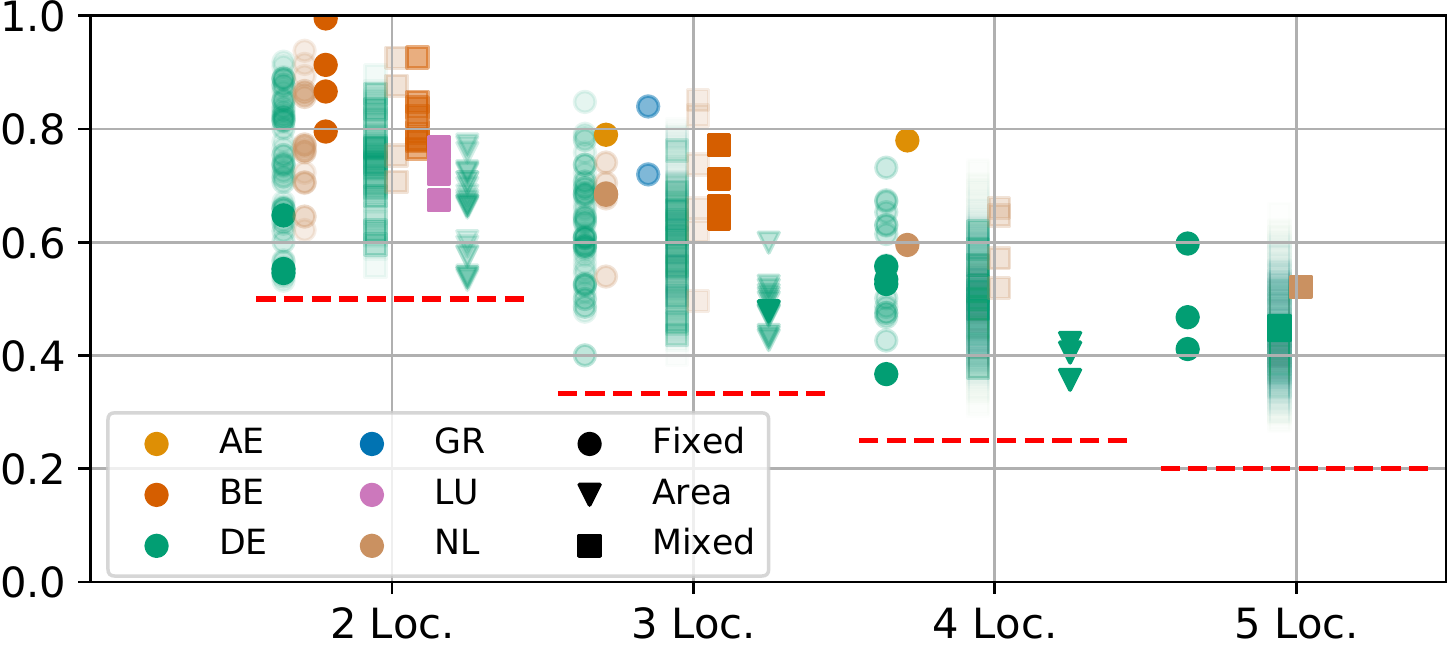}
    \caption{Classification accuracy of regional/national classifications within the same country. Dashed red lines indicate the probability of randomly guessing the correct location.}
    \label{fig:r1-summary}
\end{figure}

The performances of classifications are highly variant depending on the sets of locations. Figure~\ref{fig:r1-summary} illustrates the distribution of the performance of all classifications. 
We also present detailed results for individual classifications for all pairs of locations for each country in Tables~\ref{tab:class-results-binary-lu-be}~--~\ref{tab:any-matrix-nl} in the Appendix.

\subsection{Misclassification Errors}

In location identification, a misclassification error for an SMS measurement means that the timing pattern is matched to the wrong location, i.e., wrong pattern distribution. False results can arise due to various machine-learning (ML) factors, such as overfitting and model complexity, as well as in the form of outliers due to special network conditions. In any case, more sophisticated and motivated adversaries with more resources and ML expertise may enhance the model to improve the attack.

Country-based classifications are primarily impacted by factors such as adjacency between countries and network homogeneity (including similar operators), making it more challenging to distinguish locations. The impact of these factors can be seen for operators E and F in Table~\ref{tab:class-results-country} and in Figure~\ref{fig:boxplots-countries} for LU and NL. In fixed locations, timing similarities between locations (with the same UE and operator) can make the classification less precise due to congruent variance in network delays. It can be even more challenging when locations are very close and have similar signal conditions, such as NL-1 and NL-2 for operator G (Table~\ref{tab:any-matrix-nl}) with 62\,\% accuracy. However, this is not always the case, as in the classification of DE-3 and DE-4 for operator E (Table~\ref{tab:any-matrix-de}) which achieves 87\,\% accuracy. In addition, areas and mixed classifications can be similarly difficult to distinguish, as they combine measurements from multiple distinct locations and may overlap. Nonetheless, Tables~\ref{tab:any-matrix-nl} and~\ref{tab:any-matrix-de} include high accuracy scores even in such cases.

\section{Additional Evaluations}

In this section, we present evaluations that provide additional insights, such as the impact of geographic separation of different receiver locations, and show how an open-world classification affects the performance of the attack. We also present results from the temporal stability and network timing analyses in which we collected additional data for several DE and NL locations.

\begin{figure*}[tb]
    \centering
    \begin{subfigure}{0.45\textwidth}
        \includegraphics[width=\columnwidth]{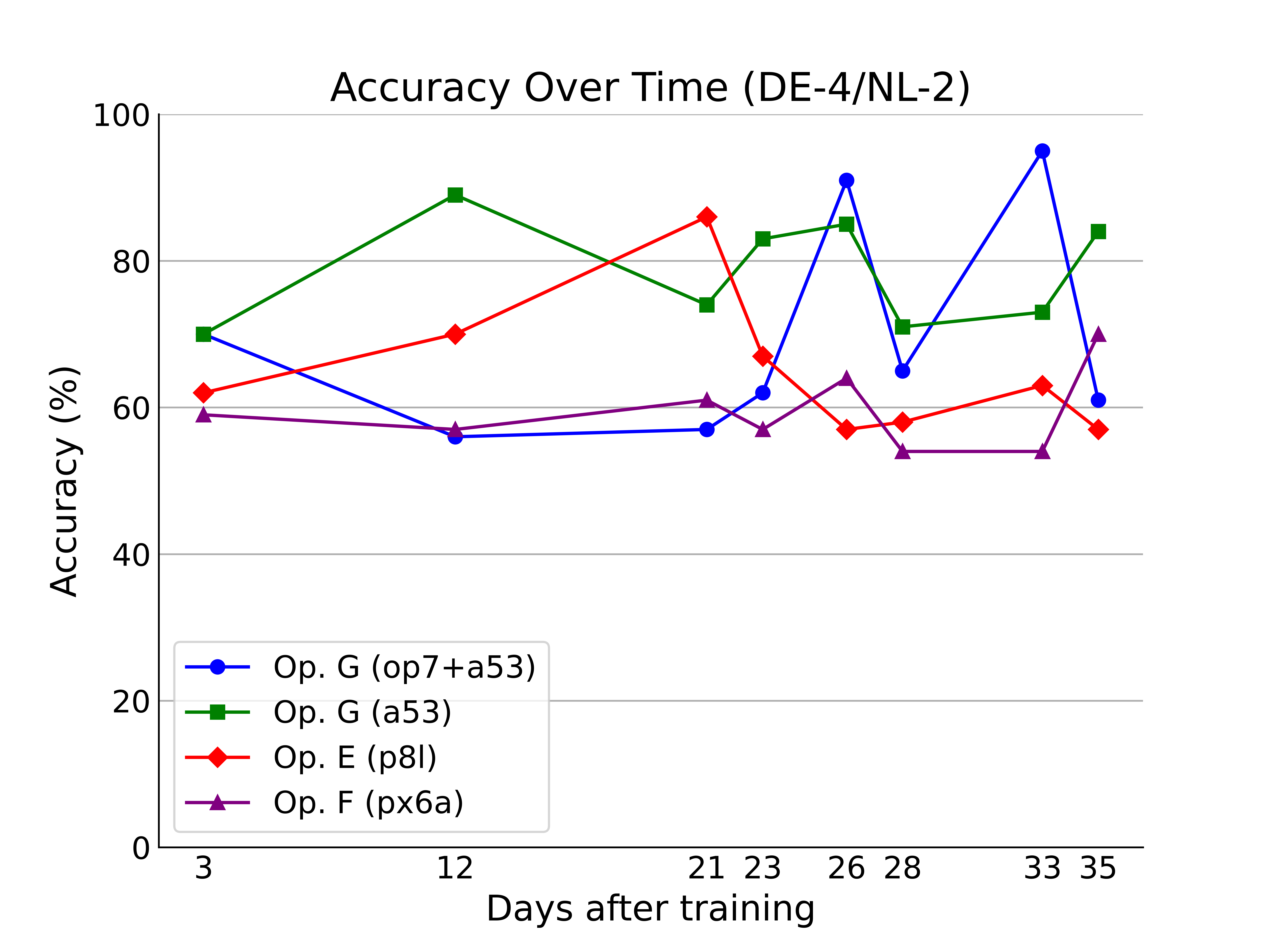}
        \vspace{-15pt}
        \caption{DE4-NL2}
        \label{fig:temp-de4-nl2}
    \end{subfigure}
    \hfill
    \begin{subfigure}{0.45\textwidth}
    \includegraphics[width=\columnwidth]{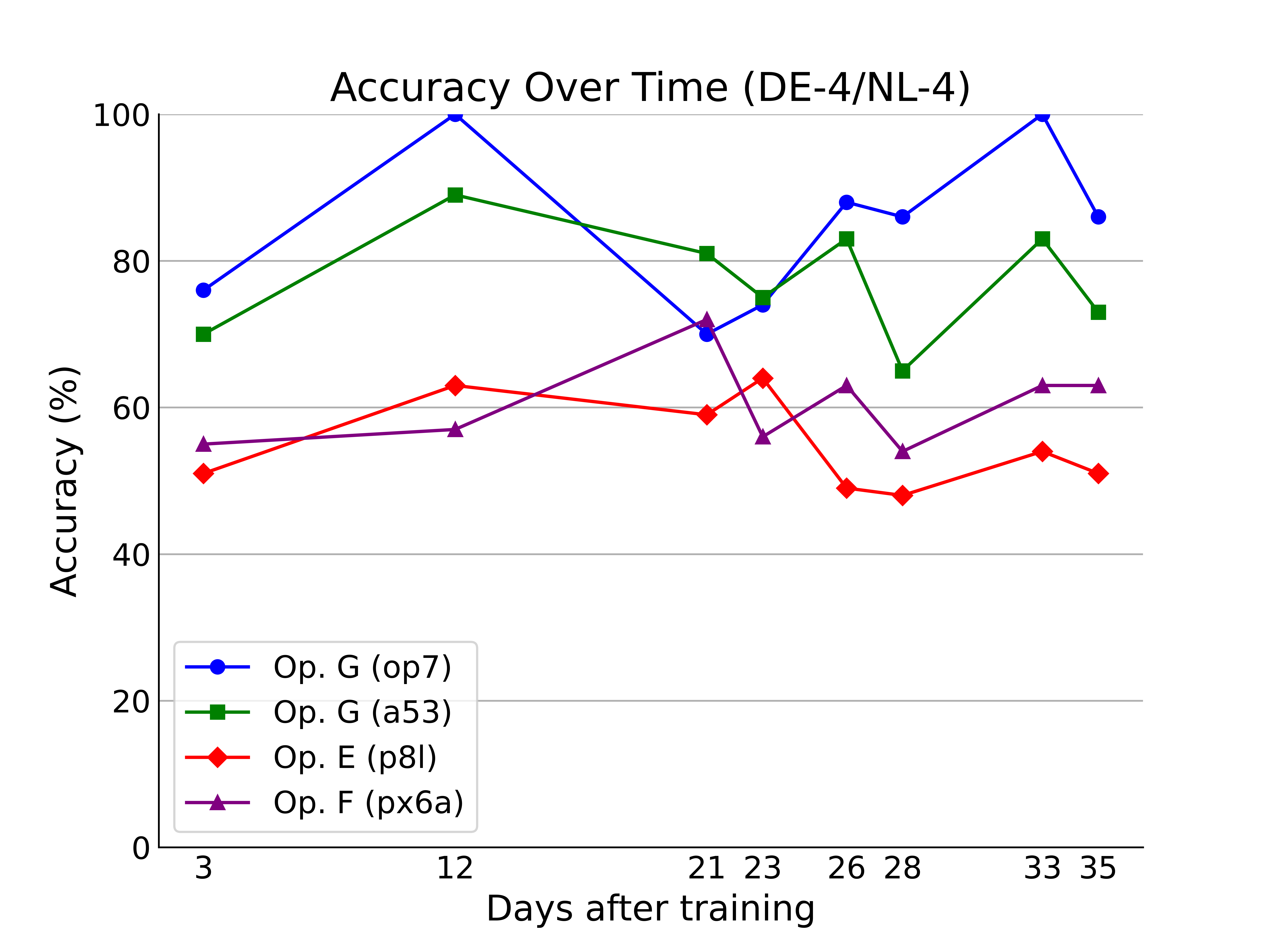}
    \vspace{-15pt}
    \caption{DE4-NL4}
    \label{fig:temp-de4-nl4}
    \end{subfigure}
    \caption{Accuracy trends of the DE4-NL2 (a) and DE4-NL4 (b) classification for various operators and devices until 35 days from the model training.}
\end{figure*}

\subsection{Temporal Stability}
\label{sec:temporal-stability}

We perform a temporal stability analysis to determine if the attack can still work even after some time has elapsed since the model was trained. For this purpose, we modified the original attack evaluation by training the model on a baseline dataset and testing it on measurements collected X days after the training phase. Therefore, we collected new and protracted data for the same locations with similar operators and devices to accommodate experimentation for up to one month and after three months from the initial training. 

Figures~\ref{fig:temp-de4-nl2} and~\ref{fig:temp-de4-nl4} depict eight examples of how the accuracy fluctuates for DE-4/NL-4 and DE-4/NL-2 classifications in a span of 35 days. We used operators G, E, and F with Huawei P8 Lite (p8l), Google Pixel 6a (px6a), Samsung Galaxy A53 (a53), and OnePlus 7 Pro (op7) devices. Each trend represents specific device(s) and operator. According to the graphs, each combination has a distinct trend, as their measurements' characteristics differ. Consequently, increases and decreases in accuracy between various days are also expected for classifications in which the model scores both with high and low accuracies. Furthermore, in Figures~\ref{fig:temp-de4-nl2} and~\ref{fig:temp-de4-nl4}, operator E with the p8l device is more susceptible to degradation than the rest of the combinations, but it takes more than 23 days for the degradation to slowly appear. Operator G with p8l in the DE-4/NL-4 classification shows a small degradation but retains high accuracy after 35 days. As a result, the collection of new data and retraining may not be necessary for all classifications. For classifications that continue to have high scores, the attacker may continue using their data. 

Figures~\ref{fig:temporal-de} and~\ref{fig:temporal-de-nl} in the Appendix show the average and maximum accuracy distributions for DE locations and the combination of DE and NL locations after three months from training. While in both figures there are classifications that still maintain high scores, several classification scores have deteriorated reaching 50\,\% or even below. This indicates that retraining for those classifications is necessary in the long term with new data in order to recalculate a baseline with high accuracy. 

\subsection{Network Analysis}
\label{sec:network-analysis}

We evaluate the impact of congestion, potential network changes, and other time-varying characteristics by running the location classification separately for different days and times of the week.
The classification process is the same as the regular attack but with specific test data slices for different times of the day and days of the week. We grouped measurements into four sets for different times of the day (0-5, 6-11, 12-17, and 18-23) and seven sets for days of the week.
We use data collected at two locations (DE-4 and NL-4) with sufficient measurements in our dataset for separate analyses across time slices, multiple phones, and operators. 

\begin{figure}[tb]
    \centering
    \includegraphics[width=\columnwidth]{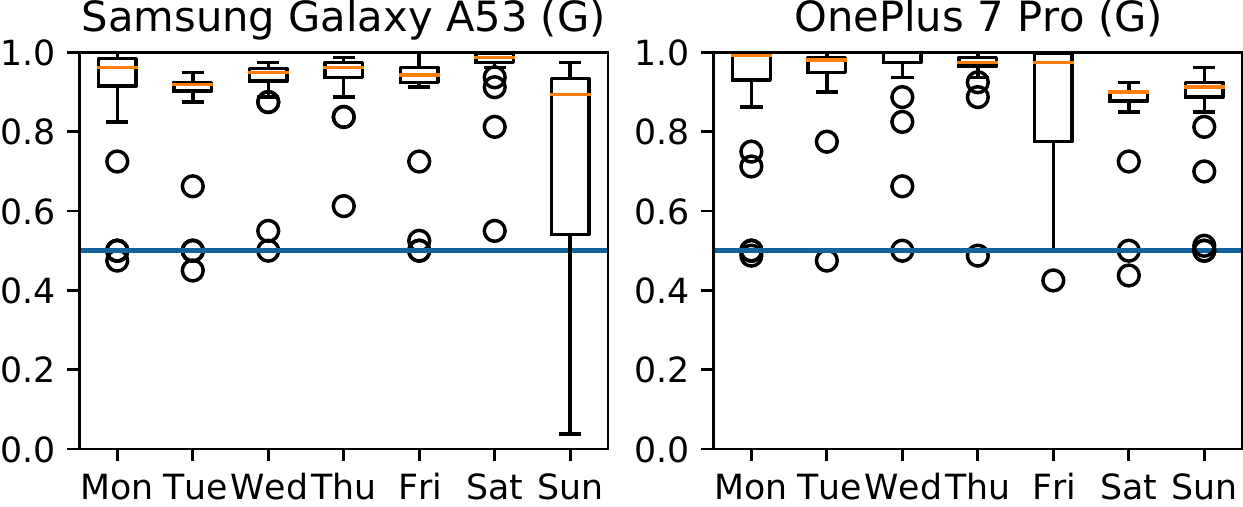}
    \caption{Network analysis of DE-4/NL-4 classification for different days and devices (Operator G).}
    \label{fig:network-wd}
\end{figure}

\begin{figure*}[tb]
    \centering
    \includegraphics[width=\textwidth]{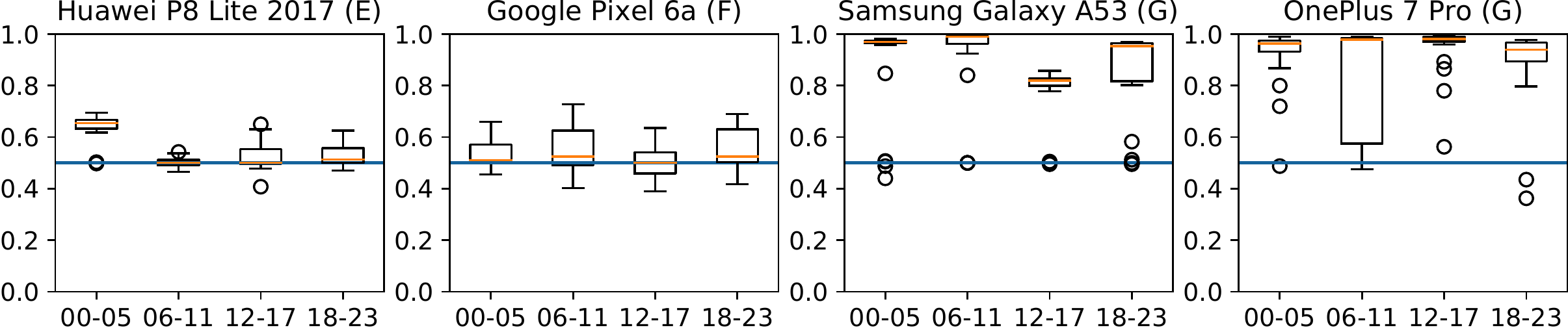}
    \caption{Network analysis of DE-4/NL-4 classification for different time windows and devices (Operator G). The figures show an example of accuracy scores for a certain combination of locations, devices, and operator.}
    \label{fig:network-hod}
\end{figure*}

Figure~\ref{fig:network-wd} shows the classification accuracy for two victim phones with one operator (G) throughout the entire week. 
The scores remained consistently high, with scores of 88\,\% and 89\,\% for OnePlus and Samsung, respectively. 
Figure~\ref{fig:network-hod} also shows the model's performance for different time windows using four phones with three different operators.  
While performance differs across operators, with classification only working for G achieving around 80\,\% and above, the scores generally remain stable throughout the day. The experiments illustrate results for specific locations, devices, and operators, and hence do not allow to draw general conclusions regarding the localization accuracy of specific devices. For the purpose of completeness, we acknowledge and report less accurate results as well.

We further evaluate the delay across different operators and locations. Figures~\ref{fig:de-times-telefonica}-\ref{fig:de-times-vodafone} show the distribution of timing delays for DE-1, DE-2, and DE-4 and operators E, F, and G, aggregated from all available phones from Figure~\ref{fig:network-hod}. We do not observe significant deviations in distributions throughout the day for those locations and conclude that network characteristics are unlikely to substantially affect the model's performance.

\subsection{Distances Between Locations}
\label{sec:eval-distances}

\begin{figure}[tb]
    \centering
    \includegraphics[width=\columnwidth,keepaspectratio]{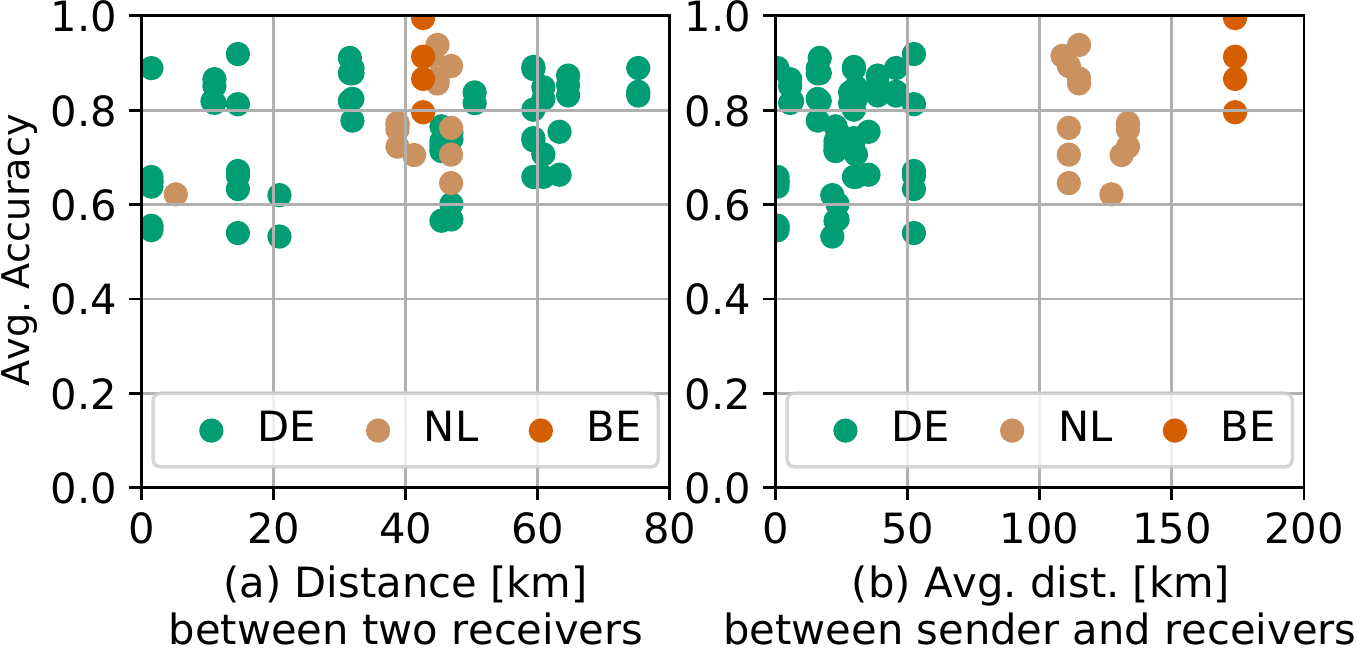}
    \caption{Accuracy of classifications with two locations, depending on (a) distances between both receiver locations and (b) between sender and receivers}
    \label{fig:accuracy-by-distance}
\end{figure}

In this part, we analyze the relationship between classification accuracy and distances between the locations using pairs of two \emph{fixed} locations in Germany, The Netherlands, and Belgium.
Figure~\ref{fig:accuracy-by-distance} shows the average classification accuracy for all pairs of locations in these three countries.
It reflects the impact on accuracy of (a) the geographical distance between the two receiver locations, and (b) the distances between the sender and each of the receiver locations.
For the latter, we consider the average of the two distances.

We found no correlation between distances and accuracies, contradicting the assumption, that receiver locations further apart from each other or from the sender would result in more accurate classification.
Therefore, distance may not be the main factor affecting classification accuracy.

\subsection{Open-world Scenarios}
\label{sec:eval-open-world}

Open-world cases refer to unknown/unseen locations, for which the attacker has not accumulated measurements for model training. We discuss three methods to tackle these cases that can be used separately or in combination.

First, the attacker can utilize outlier/anomaly detection mechanisms and unsupervised one-class classifications to reduce the "nearest neighbor" effect and identify if the data belong to an unknown location. Although this separate study requires thorough experimentation and we consider its comprehensive evaluation as future work, we carried out an experiment using an \texttt{Isolation Forest} model. The model was configured with 100 estimators (without parameter tuning) and was trained on the domestic (AE) dataset attempting to identify overseas measurements during the prediction phase. With each class having 1200 samples, as indicated in Table~\ref{tab:class-results-country}, it achieved an 88\% accuracy for anomaly detection indicating that the predicted data belong to an unseen location.

\begin{figure}[htbp]
    \centering
    \includegraphics[width=\columnwidth]{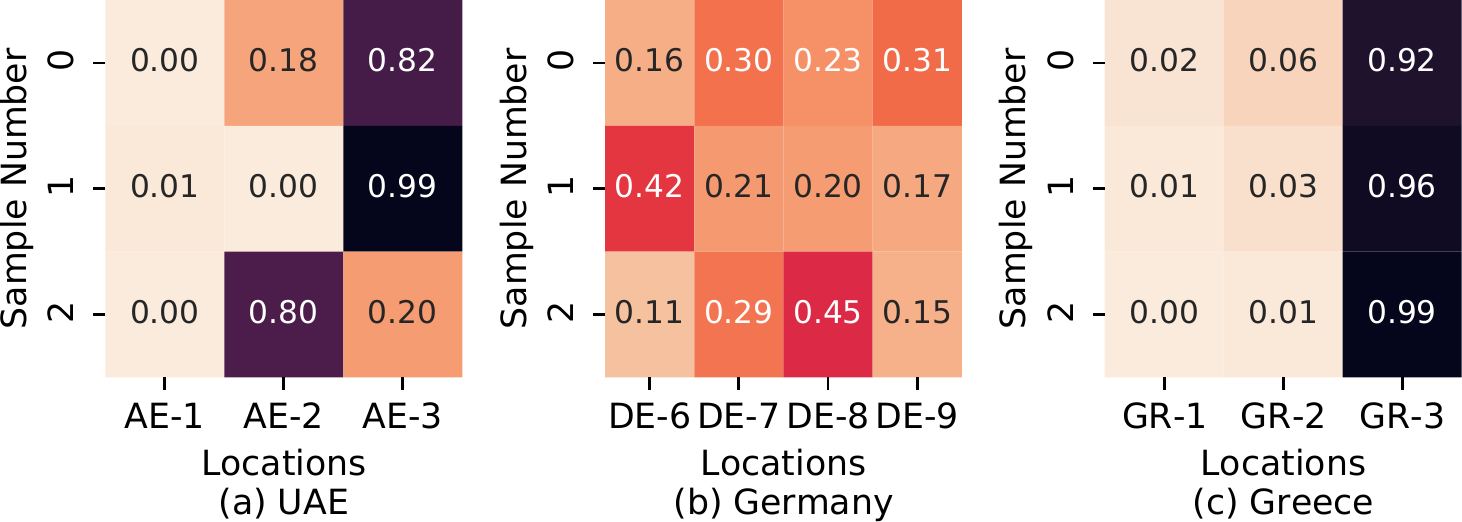}
    \caption{Probability matrices for fixed locations and areas.}
    \label{fig:heatmap-openworld}
\end{figure}

Second, the attack can be enhanced by modifying the MLP classification model to output the probability of the user being in a specific location instead of the predicted class. We have modified our initial model to run further experiments. Figure~\ref{fig:heatmap-openworld} illustrates the probabilities (per row) for fixed and area classifications in AE, DE, and GR with three distinct SMS transmissions/samples (\ie, 0, 1, 2), respectively. For AE and GR, the results show that the probabilities do not fall below 80\%. For the specific DE area classifications in Figure~\ref{fig:heatmap-openworld}b, the probabilities are more evenly allocated since the model cannot decisively decide the true class, especially in the first SMS transmission. In this case, the attacker may perform further assessments for the top two (DE-9 and DE-7) classes, or conclude that the victim might be located in one of them. 

Third, the adversary can reduce the chances of unknown classes by expanding the measurement campaign to more potential locations that are not routinely tied to the victim (\eg, famous landmarks). There are research works (focusing on WiFi) that collect data from various places within cities and areas~\cite{Schepers21:WiFi-trends, schnitzler23:hope-delivery-extracting}, while targeting either Access Points (APs) or smartphone devices. Additionally, the attacker can focus on utilizing areas instead of fixed positions to expand the coverage. Although this approach may not reveal the exact position (which can be translated to GPS coordinates) of the victim if the area incorporates too many positions, it allows the attacker to still track the victim without relying on the routinely fixed locations. However, the extensive data an attacker needs to collect beforehand may limit the practicality of this approach. In general, the attacker might prefer to resort to a binary decision, \ie, to determine whether or not the victim is at one of their previously seen locations, as described in the first two methods. In Section~\ref{further-discussion} in the Appendix, we provide more information on how the attacker can manage large dataset collections in the context of handling unseen locations. 

\section{Discussion}

Our study provides insights into how different locations of SMS receivers can be distinguished based on measuring the time it takes to deliver an SMS. In this section, we discuss potential countermeasures to mitigate the attack at different levels as well as the limitations of our study.

\subsection{Countermeasures}

\noindent\textbf{UE-based countermeasures.} On UE devices, defenses can be implemented at the application layer or become a part of the system firmware which could be suitable for low-level cellular traffic control. To our knowledge, there is no significant progress so far apart from Qualcomm's demonstration of rogue base station detection~\cite{qualcomm21:demo}. On the other hand, application-based defenses elaborate on false base station detection~\cite{Dabrowski14:IMSI-Catcher-Catchers, Park17:White-Stingray, Ney17:SeaGlass, Dabrowski16:Operator-IMSI-Catcher, Li17:FBS-Radar, Quintin21:LTE-Base-Station-Detection, Nakarmi19:Multi-RAT, arif20:Phoenix-Runtime-Verification, li2016:mobileinsight}, and on malicious SMS detection (\eg, binary, silent, \etc)~\cite{srlabs:snoopsnitch, AIMSICD, darshak}. RILDefender~\cite{Wen23:rildefender} expands the SMS attack detection by monitoring the Radio Interface Layer.

Nonetheless, we do not consider that these detection mechanisms are applicable in our case since we do not operate a false base station and do not solely rely on silent SMS. Measurement collection and prediction can happen through regular SMS as well. Therefore, there is currently no actual countermeasure against our timing attacks. Moreover, these approaches have several other drawbacks. They lack preventive countermeasures, which means that the attack has already succeeded by the time the user is potentially alerted. Furthermore, they may rely on the user to manually block potential attacks, while legitimate SMS use cases could be rejected too. Practicality is further decreased as these applications cannot be supported by devices other than Android OS and specific basebands while rooting of the device is required for the application to capture and analyze the traffic. Consequently, the only countermeasures could be to either manipulate the Delivery Reports with a random delay or not send them at all. 

\noindent\textbf{Network-based countermeasures.} Currently, no countermeasures exist to thwart location identification against a network subscriber. In fact, the network possesses neither the detection nor the prevention mechanisms to hamper or make timing attacks unattainable. However, as a first response, the operators could disable silent SMSs across their network. Although timing attacks are still feasible, the attacker will be forced to use only regular SMSes to collect measurements and interact with the victim, which is less stealthy.

In addition, operators will need to maintain a resilient spamming/flooding filter in the core network, either in the IMS or SMSC, to capture incessant transmissions destined for a specific target. The suspicious communications can either be dropped or intentionally delayed to obstruct the attack. Nevertheless, this approach may significantly impact performance for normal users. As an alternative and more holistic countermeasure, the operators could alter all SMS timings uniformly or randomly to disrupt any side-channel analysis. This could occur during the routing and processing in IMS and SMSC. Once again, this can lead to significant performance degradation which can spread to entire networks.

Finally, a draconian but effective solution would be to eliminate Delivery Reports altogether. Nonetheless, it would necessitate considerable architectural modifications in the core network and smartphone devices (\eg, baseband modems) and re-evaluation of the specifications. Additionally, it is a challenging attempt because it would require worldwide adoption and impede the user experience, network performance testing, and commercial usage (\eg, marketing).

\subsection{Limitations}
Due to our practical approach involving the same device(s) being physically placed in different geographical locations, we are limited in the amount of data that can be collected within a reasonable time period.
However, we consider that the data collected in 34 different locations spread across 10 countries provide sufficient insights to demonstrate the severity of the potential threat.

Our evaluations have demonstrated that the attack works with varying performance, depending on the sets of possible locations of a victim.
While classifications can reach high accuracy of 95\,\% and more, the attack does not work well under \emph{all} circumstances, which is an expected outcome in empirical measurements in the real world.
However, as our extensive evaluations at different granularity levels show, performance may not be a matter of distances between locations (\cf~Section~\ref{sec:eval-distances}). 
This implies that there are multiple factors contributing to the success of the attack and, thus, this may demand further in-depth evaluations and insights in future work.

While the main part of our evaluations focuses on the performance of the attack in scenarios with distinct sets of known locations of the victim (closed world), cases involving new locations the attacker does not know in advance (open world) might be even more intriguing.
Whereas we already peek into this direction in Section~\ref{sec:eval-open-world}, we argue that the attack working in \emph{closed world} scenarios already poses a significant privacy threat. 

\section{Related Work}

Most related to our research is the work by Schnitz\-ler~\etal~\cite{schnitzler23:hope-delivery-extracting} that explored the feasibility of distinguishing the location of message recipients' in messenger applications. In contrast, our SMS-based timing side-channel has more severe consequences as it relies on a fundamental and universal technology built into every mobile device in the world and cannot be mitigated. 
Our attack is not limited to specific applications and can remain undetected through silent SMSes.

Several works~\cite{Hussain19:Privacy-Attacks-Paging, Hong18:GUTI-Reallocation, Lakshmanan21:Identification-Carrier-Aggr} attempt to localize the cellular network user actively or passively by capturing identifiers. 
Shaik~\etal~\cite{Shaik2016:Practical-Attacks-LTE} (extending on~\cite{Kune12:GSM-Location-Leaks}) leverage paging messages and insecure measurement and Radio Link Failure (RLF) reports to reveal GPS locations in certain situations. 
Other localization approaches capitalize on the MAC layer and timing advance values for localization~\cite{Roth:LTE-Location-Privacy, Pimentel13:4G-Geolocation}. 
Ltrack~\cite{Kotuliak22:l-track} demonstrated improved localization to as much as 20 m with Timing Advance and more passive adversaries, and capitalizing on overshadowing techniques~\cite{Yang19:Overshadowing-Attack, Erni2021:AdaptOver}.
Lakshmanan~\etal~\cite{Lakshmanan21:Identification-Carrier-Aggr} demonstrate that sniffers collecting the activation bitmap broadcast in the public scheduling channel and the target's identifiers can identify a path taken by a target among a list of candidate paths, with a scale below 1 Km.
The adversary must accumulate sufficient measurements by taking the paths multiple times.

Our work differs in that we do not target the communication channels, use false base stations or sniffers, and are not geographically constrained. 
Deploying false base stations~\cite{Jover16:LTE-Location-Tracking-And-Exploits} may prove not only complex in many scenarios but also less stealthy, especially for active attackers that leverage malicious attachments and MitM~\cite{Rupprecht-20:Imp4gt, Bitsikas21:Vulnerabilities-Handover, Bitsikas22:5G-Emergency-System, Rupprecht-19:Layer-Two}. 
Passive attackers may require proximity to the target, especially with 5G SA, where beamforming imposes additional positioning constraints. 
Therefore, many suggested countermeasures against false base stations ranging from 3GPP studies~\cite{3gpp.33.809} to PKI mechanisms~\cite{Singla21:5G-Secure-Connection, Hussain19:Insecure-Connection} and detection techniques~\cite{Dabrowski14:IMSI-Catcher-Catchers, Park17:White-Stingray, Ney17:SeaGlass, Dabrowski16:Operator-IMSI-Catcher, Li17:FBS-Radar, Quintin21:LTE-Base-Station-Detection, Nakarmi19:Multi-RAT, arif20:Phoenix-Runtime-Verification} are ineffective.

Various SMS attacks have been demonstrated in the past, such as Simjacking~\cite{adaptivemobile19:simjacking} that exploits the vulnerabilities of the S@T Browser technology to extract sensitive user information and execute commands. \cite{Hua16:LTE-IMS-Threats} explores spamming, spoofing, DoS, and silent SMS that could impact the LTE network. 
Furthermore, Mulliner~\etal~\cite{Mulliner11:SMS-Of-Death} introduced a vulnerability analysis framework that is used to monitor unexpected smartphone behavior leading to large-scale DoS attacks. 
Apart from SMS, audio call features have been explored~\cite{Peeters18:Sonar, Balasubramaniyan10:PinDr0p} for fingerprinting and anomaly detection to detect call redirection/hijacking. Sonar~\cite{Peeters18:Sonar} uses the audio latency with the round-trip time due to distance, while PinDr0p~\cite{Balasubramaniyan10:PinDr0p} leverages the applied codes, packet loss profiles and bit error rates.

\section{Conclusion}

In this work, we introduced a novel timing side-channel attack for exploiting the SMS procedure, allowing us to distinguish between receivers in locations within a specific region, country, or abroad. We have demonstrated that the SMS procedure leaks timing delays related to the receiver and operator, and we constructed an attack that uses silent SMS to remain stealthy. In addition, we argue that the attack can reach every user who possesses a smartphone device and is subscribed to a network operator. This increases the impact and practicality of the attack as the adversary needs only the victim's phone number to collect measurements from their usual locations of the target. Finally, we clarify that it is hard to enforce countermeasures against timing attacks due to the required architectural modifications, SMS worldwide use, and performance overhead.

\section*{Acknowledgements}
This work was supported by NSF grant 2144914, by UA Ruhr under the Research Alliance Ruhr program, by the Abu Dhabi Award for Research Excellence 2019 (\#AARE19-236), and by the Center for Cyber Security at New York University Abu Dhabi (NYUAD). The authors would like to thank Michel Lang, Philipp Markert, and Lena Schnitzler for their help with data collection.

{\footnotesize \bibliographystyle{plain}
\bibliography{bibliography.bib}}

\appendix

\section{Neural Network Parameter Tuning} \label{tuning}

We used manual and automatic parameter tuning. For manual tuning, we mainly experimented with the neural network layers. For automatic parameter tuning we explored the following various setups:

Parameters: [ \{ \\
\tabto{36pt}    \textbf{hidden\_layer\_sizes:} (10,40,10), (8,8,8), \\ \tabto{42pt} (10,10,10), (8,10,8), (10,50,10), (10,60,10) \\
\tabto{36pt}    \textbf{activation:} (tanh, relu, logistic, identity) \\
\tabto{36pt}    \textbf{solver:} (sgd, adam) \\
\tabto{36pt}    \textbf{alpha:} (0.0001, 0.001, 0.005)  \\
\tabto{36pt}    \textbf{learning\_rate:} (constant, adaptive) \\
\tabto{36pt}    \textbf{max\_iter:} (100, 200, 500, 1000, 2000, 5000) \\
\tabto{36pt}    \textbf{momentum:} (0.2, 0.5, 0.7, 0.9)\\
\tabto{26pt}      \},]

\section{Further Discussion}\label{further-discussion}

\noindent\textbf{Can the scalability and feasibility be improved?} The process can be automated (including SMS exchange, measurement collection, data processing, training, and prediction) with minimum requirements in terms of equipment. The attacker does not have to be constantly involved in the process, apart from occasional monitoring. Additionally, in our work, we perform this attack with a single device-sender at a time from one location, while multiple devices at the same time could be deployed to expedite the process. It is also possible to increase the SMS exchange rate. Furthermore, the dataset size is not a hurdle for the location identification attack, since silent SMS can be stealthily utilized and the attack works with small and large datasets (Table~\ref{tab:counts}, i.e., 350-15,000 SMS). Hence, the attacker may choose to collect a smaller amount of data for training and prediction. 

\noindent\textbf{What if the SMS Delivery Report fails?} During the experimentation we very rarely noticed that a Delivery Report failed. As mentioned in Section~\ref{measurement-collection}, we defined the SMS burst accordingly in order to avoid network congestion, flooding and errors. Furthermore, the slow rates increase the stealthiness of the attack. In case a Delivery Report fails, we ceased the transmission for 10 minutes and then continued without any issue until we collected the complete dataset.

\noindent\textbf{What if the victim is using the device at the moment of the attack?} Cellular services typically include calls, SMSs, Internet access (\eg, web browsing, video streaming, social applications, \etc). These activities, especially SIP calls, may culminate in fluctuations in the timings, if the channel is heavily occupied. The attacker can distinguish and filter out such timings since they tend to be statistical outliers.

\begin{figure}[ht]
    \centering
    \includegraphics[width=\columnwidth,keepaspectratio]{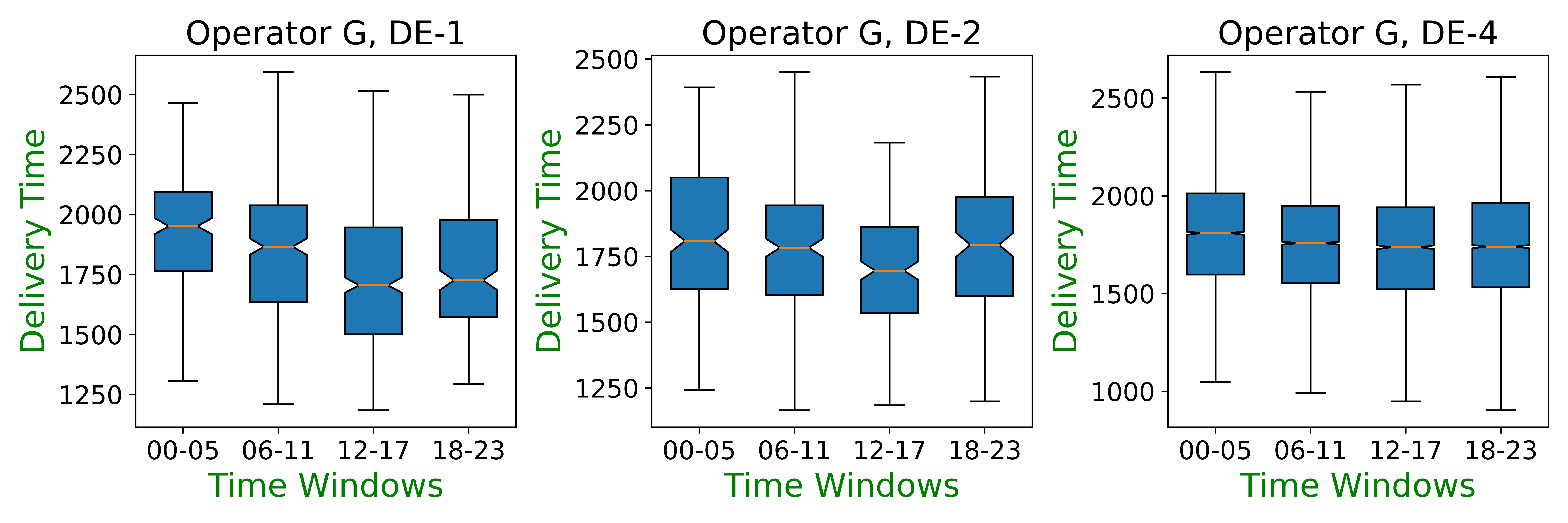}
    \caption{Delivery time plots based on different times of the day for DE locations (Operator G).}
    \label{fig:de-times-telefonica}
\end{figure}

\begin{figure}[ht]
    \centering
    \includegraphics[width=\columnwidth,keepaspectratio]{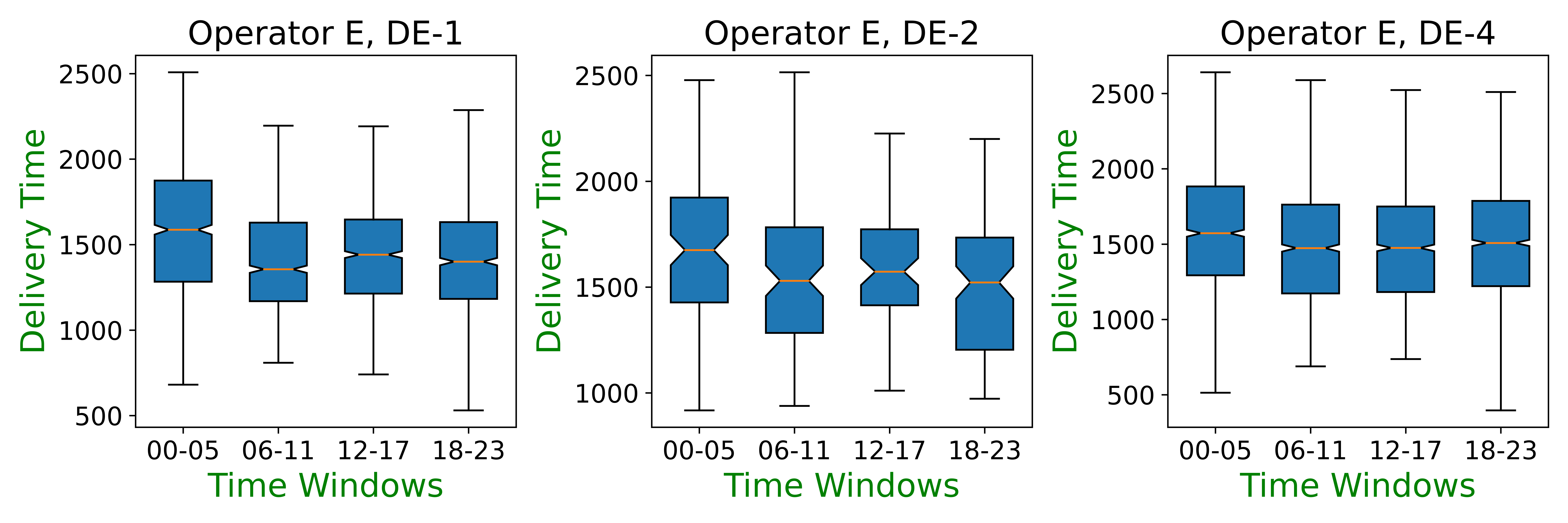}
    \caption{Delivery time plots based on different times of the day for DE locations (Operator E).}
    \label{fig:de-times-telekom}
\end{figure}

\begin{figure}[ht]
    \centering
    \includegraphics[width=\columnwidth,keepaspectratio]{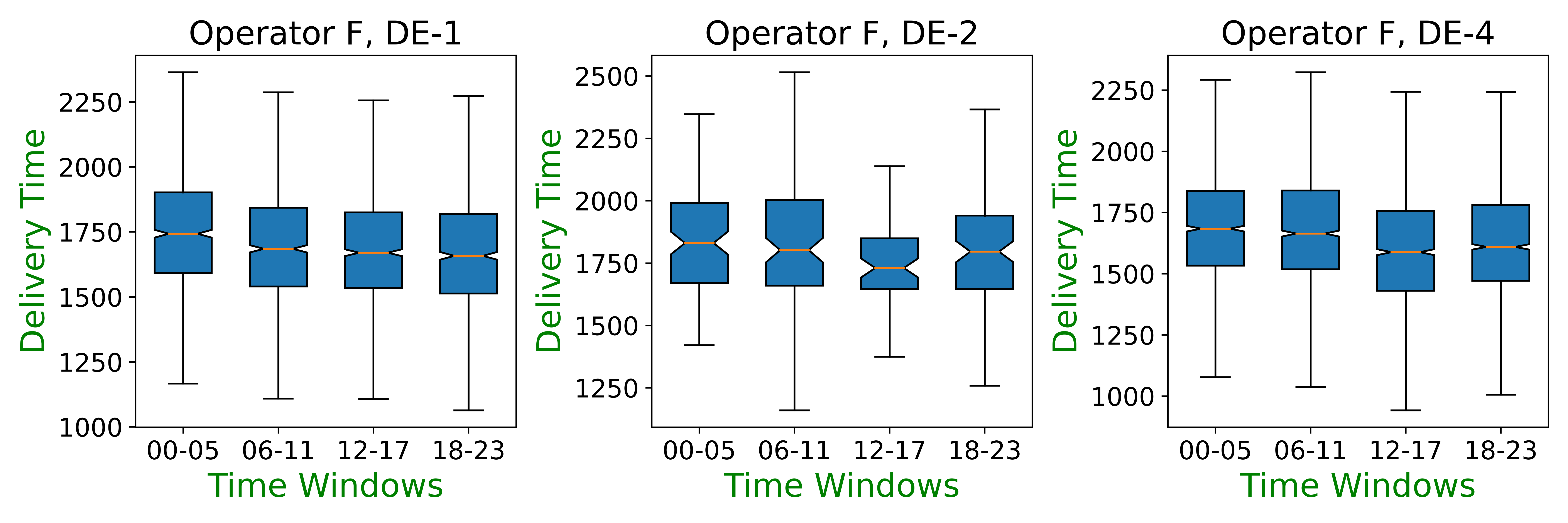}
    \caption{Delivery time plots based on different times of the day for DE locations (Operator F).}
    \label{fig:de-times-vodafone}
\end{figure}

\begin{figure*}[t!]
\begin{subfigure}[t]{\columnwidth}
    \centering
    \includegraphics[width=\columnwidth]{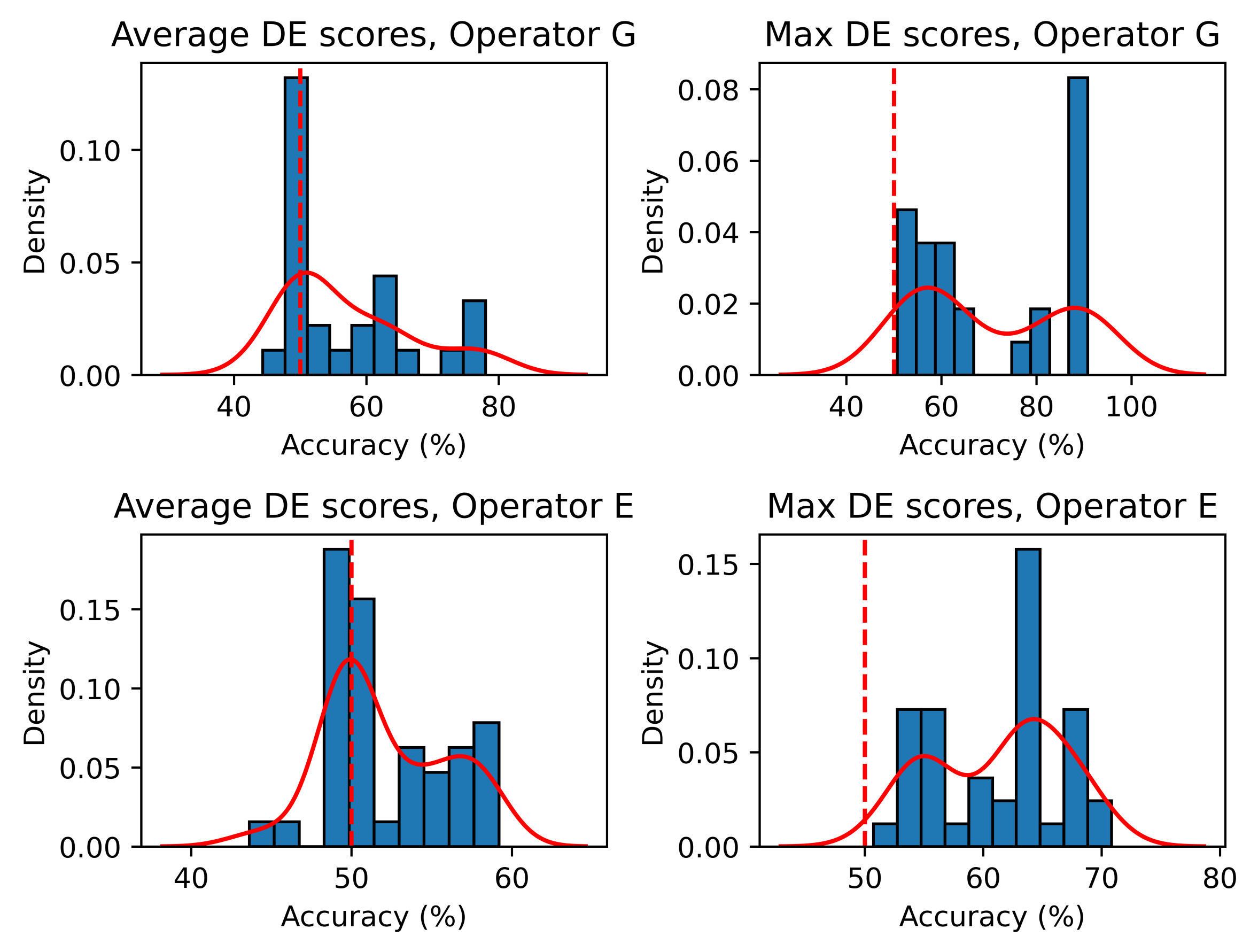}
    \caption{Accuracy distributions for DE locations.}
    \label{fig:temporal-de}
\end{subfigure}
\hfill
\begin{subfigure}[t]{\columnwidth}
    \centering
    \includegraphics[width=\columnwidth]{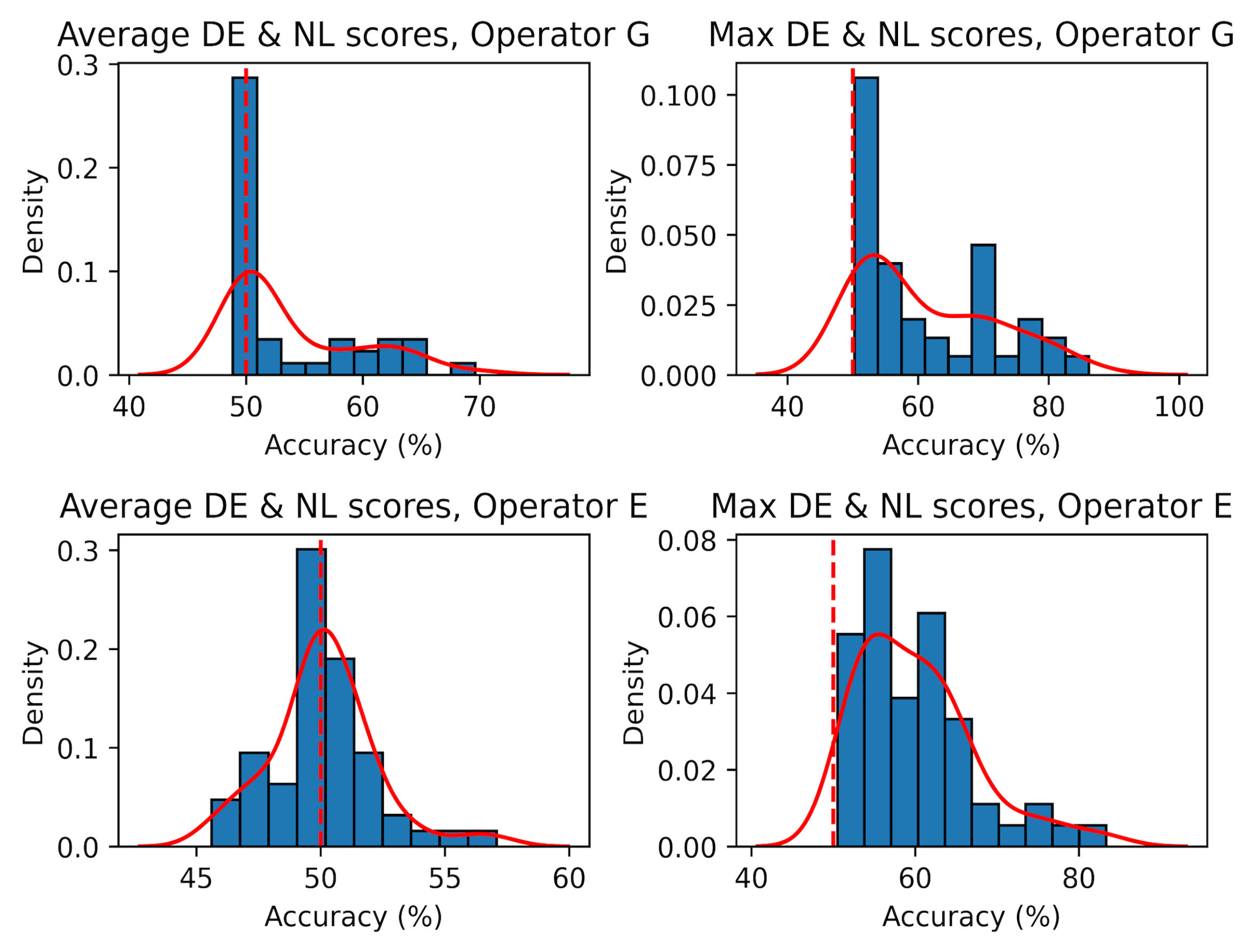}
    \caption{Accuracy distributions for DE and NL locations.}
    \label{fig:temporal-de-nl}
\end{subfigure}
\caption{The graphs illustrate the distributions of all classifications for (a) DE and (b) DE with NL locations. The prediction/attack phase considers measurements collected by one sender device for operators G and E \textbf{three months after} the initial model training.}
\end{figure*}

\subsection{SMS Timings Across Operators}

In this section, we conduct classifications for various operators, \ie, we distinguish timings measured for different operators. The goal of this classification is to identify diversities in networks that affect SMS timings. Table~\ref{tab:class-results-operator} presents our experimental results.

\begin{figure}[tb]
    \centering
    \includegraphics[width=\columnwidth]{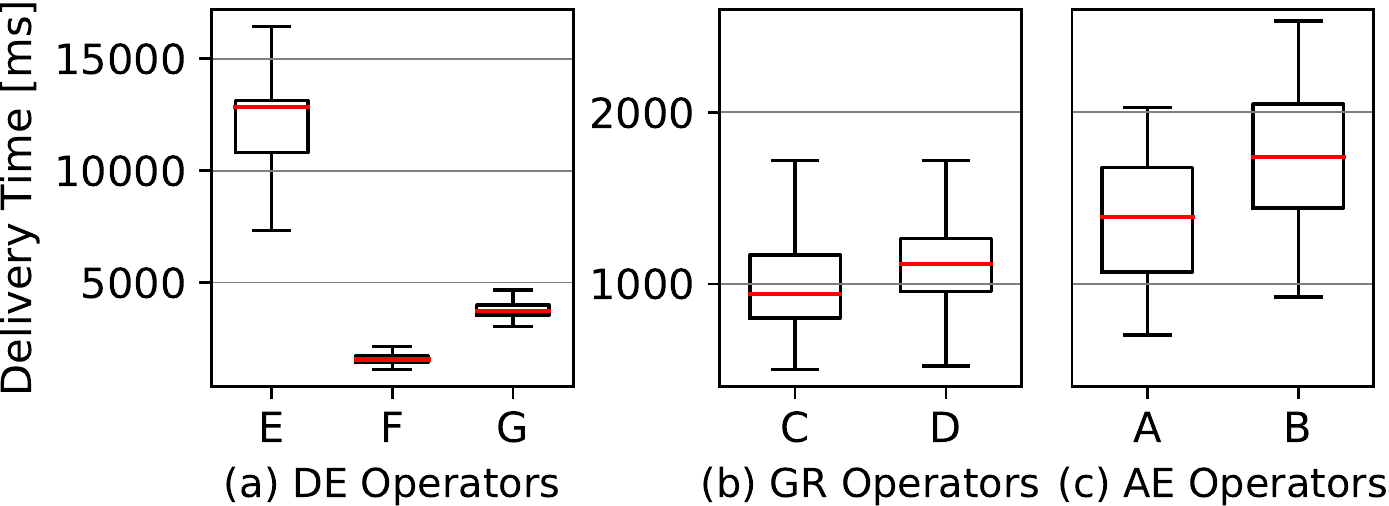}
    \caption{Delivery timings across different operators in various countries.}
    \label{fig:boxplots-operators}
\end{figure}

We perform experiments in AE, GR, DE, NL, BE, and LU for different locations and operators. We achieve at least 99\% average accuracy for AE-2 with operators A and B, and 82\% accuracy for GR-1 with operators C and D. Similarly, DE locations achieve at least 88\% accuracy for E, F, and G. The dataset size for all training and prediction procedures ranges from 220 to 578 timing signatures. The plots in Figure~\ref{fig:boxplots-operators} clearly show the distinctness of delivery timings between operators in DE, GR, and AE.

\noindent\textbf{Roaming cases.} We analyze the remaining cases separately due to the inclusion of measurements from roaming connections. The highest accuracy is achieved by AE-1 classification for A and C operators (Table~\ref{tab:class-results-operator}). In contrast, neighboring countries such as NL, BE, and LU show less heterogeneity, with LU-1 achieving an average accuracy of 69\%, and BE-1 and BE-2 reaching 61\% and 66\%, respectively. NL-1 and NL-4 produce scores that do not exceed 52\% in our setup. Kindly note that random guessing is 33\,\% with three classes.

\begin{table}[!b]
\centering
\caption{\label{tab:class-results-operator} Classification results for various operator.}
\begin{tabular}{@{}clcc}
\toprule
\textbf{Samples} & \textbf{Operators} & \textbf{Rec. Loc.} & \textbf{Accuracy}\\ 
\midrule
\multicolumn{4}{l}{\emph{Sender Location: AE-1}}\\
220 & A,C & AE-1  & 100\% \\
220 & A,B & AE-2 & 99\% \\
300 & A,B & AE-3  & 100\% \\
\midrule
\multicolumn{4}{l}{\emph{Sender Location: GR-1}}\\
300 & C,D & GR-1  & 82\% \\
\midrule
\multicolumn{4}{l}{\emph{Sender Location: DE-4}}\\
511 & E,F,G & DE-3  & 94\% \\
578 & E,F,G & DE-4  & 88\% \\
889 & E,F,G & NL-4  & 51\% \\
280 & E,F,G & NL-1  & 52\% \\
313 & E,F,G & BE-1  & 61\% \\
338 & E,F,G & BE-2  & 66\% \\
257 & E,F,G & LU-1  & 69\% \\
\bottomrule
\end{tabular}
\end{table}

\section{SMS Timings Across Devices}

We perform device classifications to determine if we can distinguish between devices demonstrating that UE processing (\cf~Figure~\ref{fig:sys-delays}) is involved in timing measurements. UE processing incorporates baseband, OS, and SIM characteristics in the timings. 

We used six smartphones from Table~\ref{tab:specs} to conduct the experiments. We deployed them at the same location with the same operator to ensure that the timings are associated with the smartphone device and cannot be influenced by the location's or network's properties. Table~\ref{tab:devices-results} separates the results into two sections and depicting the accuracy scores for GR and DE locations. 

The first experiment was conducted in GR-1 with iPhone 6 and iPhone 7 with operator C. We collected 300 SMS measurements for each device, and the results show device identification with an accuracy of 87\%. Similarly, the second part was carried out in DE-4, where operator G was used for the connections for the Google Pixel, OnePlus 7, Nokia 5.3, and Huawei P8 devices. Our dataset sizes for this part were larger than the first part with 564 SMS measurements for Oneplus-Huawei, Nokia-Huawei, and Google-Huawei comparisons, 578 for Google-OnePlus, Google-Nokia comparisons, and finally 754 for OnePlus-Nokia comparison. Results show that we can identify smartphone devices with at least 99\% accuracy in some cases, apart from the Huawei P8-Google Pixel and Nokia 5.3-Oneplus 7 classifications which present less diversity.

\begin{table}[tb]
    \centering
    \caption{Device classification results for different devices.}
    \label{tab:devices-results}
    \setlength{\tabcolsep}{1.5pt}
    \resizebox{\columnwidth}{!}{
    \begin{tabular}{@{}l|cccccc}
    \toprule
    \textbf{Devices}
    & iPhone 12 & iPhone 7 & Google Pixel & OnePlus 7 & Nokia 5.3 & Huawei P8 \\
    \midrule
    \multicolumn{7}{l}{\emph{Sender Location: GR-1}}\\
    iPhone 12 & --- & 87\% & --- & --- & --- & ---\\
    iPhone 7 & 87\% & --- & --- & --- & --- & ---\\ 
    \midrule
    \multicolumn{7}{l}{\emph{Sender Location: DE-4}}\\
    Google Pixel & --- & --- & --- & 100\% & 99\% & 66\% \\
    Oneplus 7 & --- & --- & 100\% & --- & 53\% & 100\% \\
    Nokia 5.3 & --- & --- & 99\% & 53\% & --- & 99\% \\
    Huawei P8 & --- & --- & 66\% & 100\% & 99\% & ---\\
    \bottomrule
    \end{tabular}}
\end{table}

\section{Android baseband logging} \label{baseband-log}

In our SMS Handler application, we defined the structure of the silent SMS in the Android SmsManager as follows:

\begin{lstlisting}
final byte[] payload = new byte[]{0x0A, 0x06, 0x03, (byte) 0xB0, (byte) 0xAF, (byte) 0x82, 0x03, 0x06, 0x6A, 0x00, 0x05};
\end{lstlisting}

Next, through ADB and Logcat we were able to realize the SMS procedure at the lower layers. The command which was running during the SMS transmission was: \texttt{adb logcat -b radio > radio.txt}.

\begin{figure}[!ht]
    \centering
    \includegraphics[width=1.0\columnwidth,keepaspectratio]{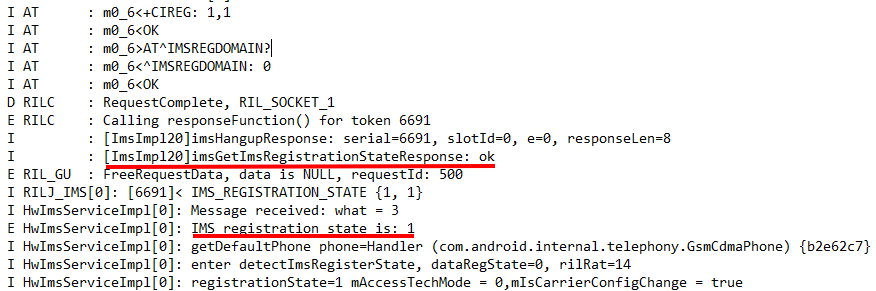}
    \caption{IMS Registration logs and AT commands}
    \label{fig:ims_reg_logs}
\end{figure}

\begin{figure}[!ht]
    \centering
    \includegraphics[width=1.0\columnwidth,keepaspectratio]{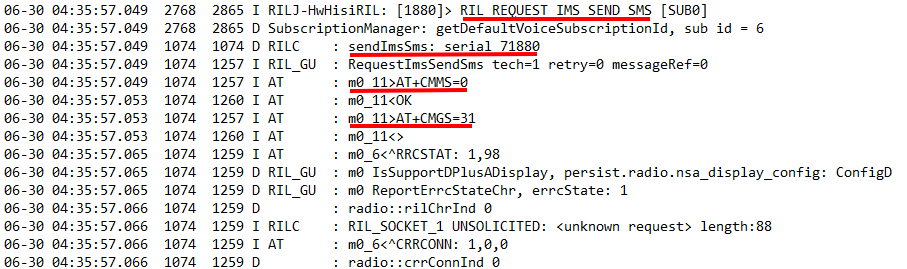}
    \caption{Send SMS logs and AT commands}
    \label{fig:send_sms_logs}
\end{figure}

\begin{figure}[!ht]
    \centering
    \includegraphics[width=1.0\columnwidth,keepaspectratio]{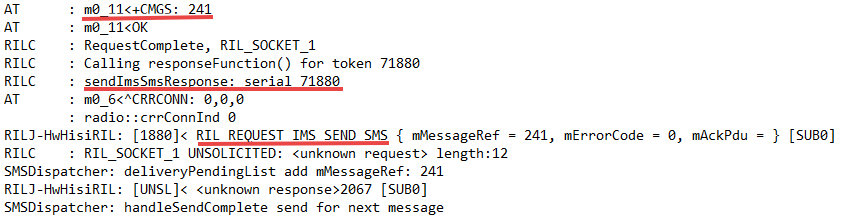}
    \caption{Acknowledgement logs and AT commands}
    \label{fig:ack_logs}
\end{figure}

\begin{figure}[!ht]
    \centering
    \includegraphics[width=1.0\columnwidth,keepaspectratio]{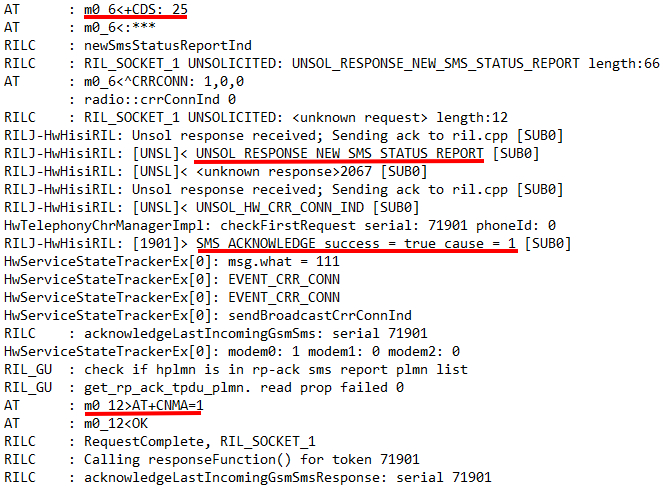}
    \caption{Status Report logs and AT commands}
    \label{fig:status_report_logs}
\end{figure}

By investigating the AT commands~\cite{3gpp.27.005} the attacker can also collect indications about the kind of connection that is used for the SMS transmissions. Figure~\ref{fig:ims_reg_logs} shows the IMS registration state prior to sending the SMS texts which in this case is enabled. Figure~\ref{fig:send_sms_logs} presents an active device that sends an SMS through IMS (with LTE). The \textit{AT+CMMS} command is used to inform the modem that several SMS messages will be sent in quick succession, and the link should be held open for more efficient transmission, but in this case it is 0 (\ie, disabled). The AT command \textit{AT+CMGS} is used to send the actual SMS text and the modem responds with the message ID, \ie, 31. Then, Figure~\ref{fig:ack_logs} illustrates the successful delivery of the SMS text ("sent" notification) and the wait-list for the Delivery Report. Finally, Figure~\ref{fig:status_report_logs} depicts the reception of the Delivery Report, where \textit{AT+CDS} notifies us of an unsolicited delivery status in Protocol Data Unit (PDU) mode. The device responds back with an acknowledgement to Delivery Report with \textit{AT+CNMA = 1}, where 1 indicates the RP-ACK.

\section{SMSoNAS vs.~SMSoIP} \label{nas-vs-ip}
 
Compared to SMSoIP, SMSoNAS has a different routing path (\cf~Figure~\ref{fig:arch}) and additional procedures that include different encryption/decryption and integrity validation processes~\cite{3gpp.24.301, 3gpp.24.501}. IP-based communications including SIP rely on outsourcing mechanisms for protection, even though the IMS AKA authenticates the subscriber. These are the IPsec and TLS which can encapsulate the payload in the network and over the transport layers respectively. On the contrary, SMS over NAS benefits from NAS layer protection without extra security enhancements~\cite{3gpp.33.501, 3gpp.24.501}. Consequently, these differences may cause divergent delays in the network which an adversary can capitalize on.

\begin{table}[!htb]
\centering
\caption{\label{tab:class-results-binary-lu-be} Classification accuracy for pairs of locations in Belgium and Luxembourg.}
\begin{tabular}{@{}lc@{}}
\hline
\textbf{Receiver Locations} & \textbf{Accuracy} \\
\midrule
\multicolumn{2}{l}{\emph{Sender Location: DE-4, Operator E}}\\
BE-1, BE-2 & 83\,\%\\
BE-1, BE-3 & 80\,\%\\
BE-2, BE-3 & 74\,\%\\
LU-1, LU-3 & 64\,\%\\
\midrule
\multicolumn{2}{l}{\emph{Sender Location: DE-4, Operator F}}\\
BE-1, BE-2 & 95\,\%\\
BE-1, BE-3 & 72\,\%\\
BE-2, BE-3 & 80\,\%\\
LU-1, LU-3 & 66\,\%\\
\midrule
\multicolumn{2}{l}{\emph{Sender Location: DE-4, Operator G}}\\
BE-1, BE-2 & 86\,\%\\
BE-1, BE-3 & 84\,\%\\
BE-2, BE-3 & 84\,\%\\
LU-1, LU-3 & 72\,\%\\
\bottomrule
\end{tabular}
\end{table}

\begin{table}[!htb]
\centering
\caption{\label{tab:class-results-multi-fixed} Multi-class classification tasks between fixed positions for AE and GR.}
\resizebox{.85\columnwidth}{!}{%
\begin{tabular}{@{}clc@{}}
\toprule
\textbf{Samples} & \textbf{Receiver Locations} & \textbf{Accuracy} \\ 
\midrule
\multicolumn{3}{l}{\emph{Sender Location: AE-1, Operator: A}}\\
 300 & AE-1, AE-2, AE-3, AE-4 & 58\% \\
 300 & AE-1, AE-2, AE-3 & 76\% \\
 \midrule
 \multicolumn{3}{l}{\emph{Sender Location: GR-1, Operator: C}}\\
 300 & GR-4, GR-5, GR-6 & 76\% \\ 
 300 & GR-1, GR-2, GR-3 & 82\% \\ 
\bottomrule

\end{tabular}
}
\end{table}


\begin{table}[th]
\centering
\caption{\label{tab:any-matrix-nl} Classification accuracy for pairs of fixed locations and areas in the Netherlands.}
\begin{tabular}{@{}lccc|c@{}}
\toprule

& NL-2 & NL-3 & NL-4 & NL-5 \\

\midrule
\multicolumn{4}{l}{\emph{Sender Location: DE-4, Operator: E}}\\

NL-1 & & 60\,\% & 52\,\% &  \\
NL-2 & -- &  &  &  \\
NL-3 & -- & -- & 52\,\% &  \\
NL-4 & -- & -- & -- &  \\

\midrule
\multicolumn{4}{l}{\emph{Sender Location: DE-4, Operator: F}}\\

NL-1 & & 50\,\% & 48\,\% & \\
NL-2 & -- &  &  &  \\
NL-3 & -- & -- & 54\,\% & \\
NL-4 & -- & -- & -- & \\

\midrule
\multicolumn{4}{l}{\emph{Sender Location: DE-4, Operator: G}}\\

NL-1 & 62\,\% & 92\,\% & 49\,\% & 68\,\% \\
NL-2 & -- & 98\,\% & 61\,\% & 58\,\%  \\
NL-3 & -- & -- & 88\,\% & 88\,\% \\
NL-4 & -- & -- & -- & 70\,\% \\

\bottomrule
\end{tabular}
\end{table}

\begin{table*}[htbp]
\centering
\caption{\label{tab:any-matrix-de} Classification accuracy for pairs of fixed locations and areas in Germany.}
\begin{tabular}{@{}lccccc|cccc@{}}
\toprule

& DE-2 & DE-3 & DE-4 & DE-5 & DE-10 & DE-6 & DE-7 & DE-8 & DE-9\\

\midrule
\multicolumn{6}{l}{\emph{Sender Location: DE-4, Operator: E}}\\

DE-1 & 74\,\% & & 63\,\% & & 79\,\% & 63\,\% & & &\\
DE-2 & -- & 77\,\% & 62\,\% & 74\,\% & 65\,\% & 68\,\% & 73\,\% & 60\,\% & 62\,\% \\
DE-3 & -- & -- & 87\,\% & 76\,\% & 72\,\% & 86\,\% & 44\,\% & 54\,\% & 62\,\% \\
DE-4 & -- & -- & -- & 75\,\% & 67\,\% & 55\,\% & 53\,\% & 64\,\% & 57\,\%\\
DE-5 & -- & -- & -- & -- & 72\,\% & 74\,\% & 73\,\% & 77\,\% & 63\,\%\\
DE-10 & -- & -- & -- & -- & -- & 64\,\% & 61\,\% & 56\,\% & 66\,\%\\
\cmidrule{2-10}
DE-6 & -- & -- & -- & -- & -- & -- & 57\,\% & 62\,\% & 56\,\% \\
DE-7 & -- & -- & -- & -- & -- & -- & -- & 63\,\% & 58\,\% \\
DE-8 & -- & -- & -- & -- & -- & -- & -- & -- & 46\,\% \\

\midrule
\multicolumn{6}{l}{\emph{Sender Location: DE-4, Operator: F}}\\

DE-1 & & 82\,\% & 58\,\%  & 56\,\% & & & 74\,\% & 74\,\% & 60\,\%\\
DE-2 & -- & & & & & & & &\\
DE-3 & -- & -- & 67\,\% & 76\,\% & & & 60\,\% & 52\,\% & 66\,\% \\
DE-4 & -- & -- & -- & 64\,\% & & & 67\,\% & 62\,\% & 52\,\%\\
DE-5 & -- & -- & -- & -- & & & 70\,\% & 82\,\% & 68\,\%\\
DE-10 & -- & -- & -- & -- & -- & & & &\\
\cmidrule{2-10}
DE-6 & -- & -- & -- & -- & -- & -- & & &\\
DE-7 & -- & -- & -- & -- & -- & -- & -- & 62\,\% & 54\,\%\\
DE-8 & -- & -- & -- & -- & -- & -- & -- & -- & 51\,\% \\

\midrule
\multicolumn{6}{l}{\emph{Sender Location: DE-4, Operator: G}}\\

DE-1 & 81\,\% & & 78\,\% & 50\,\% & 86\,\% & 74\,\% & 70\,\% & 70\,\% & 68\,\% \\
DE-2 & -- & 62\,\% & 56\,\% & 82\,\% & 91\,\%  & 64\,\% & 75\,\% & 88\,\% & 78\,\%\\
DE-3 & -- & -- & 61\,\% & 92\,\% & 82\,\% & 52\,\% & 61\,\% & 77\,\% & 72\,\%\\
DE-4 & -- & -- & -- & 82\,\% & 84\,\% & 68\,\% & 50\,\% & 84\,\% & 74\,\%\\
DE-5 & -- & -- & -- & -- & 86\,\% & 81\,\% & 72\,\% & 81\,\% & 70\,\%\\
DE-10 & -- & -- & -- & -- & -- & 76\,\% & 83\,\% & 88\,\% & 84\,\%\\
\cmidrule{2-10}
DE-6 & -- & -- & -- & -- & -- & -- & 58\,\% & 72\,\% & 58\,\% \\
DE-7 & -- & -- & -- & -- & -- & -- & -- & 58\,\% & 62\,\% \\
DE-8 & -- & -- & -- & -- & -- & -- & -- & -- & 52\,\% \\

\bottomrule
\end{tabular}
\end{table*}

\begin{table*}[bth]
\centering
\small
\caption{\label{tab:specs} Device Specifications. Except from Google Pixel 4 XL which used eSIM, all devices were equipped with physical SIM cards. The attack worked on all tested smartphones. }
\resizebox{0.8\textwidth}{!}{%
\begin{tabular}{llllc}
\toprule
\textbf{Device} & Modem  & OS & Model & Release \\ 
\midrule

\textbf{Apple iPhone 13} & Qualcomm Snapdragon X60 & iOS 15 & A2633 & 2021 \\
\rowcolor{gray!25}
\textbf{One Plus Nord 2 5G} & MediaTek Dimensity 1200 5G & Android 11 & DN2101 & 2021 \\
\textbf{Alcatel 1S} & Spreadtrum UNISOC SC9863 & Android 11 & 6025D & 2021 \\
\rowcolor{gray!25}
\textbf{Apple iPhone 12 mini} & Qualcomm X55 modem & iOS 15 & A2399 & 2020 \\ 
\textbf{Nokia 8.3 5G} & Snapdragon 765G 5G & Android 10 & TA-1243 & 2020 \\
\rowcolor{gray!25}
\textbf{Apple iPhone 12} & Qualcomm Snapdragon X55 & iOS 15 & A2403 & 2020 \\
\textbf{Samsung Galaxy A21S} & Samsung Exynos 850 & Android 10 & SM-A217F & 2020 \\
\rowcolor{gray!25}
\textbf{Huawei P40 Pro 5G$^\ast$} & HiSilicon Kirin 990 5G & Android 10 & ELS-NX9 & 2020 \\ 
\textbf{Nokia 5.3} & Qualcomm Snapdragon 665 & Android 11 & TA-1234 & 2020 \\
\rowcolor{gray!25}
\textbf{Google Pixel 4 XL} & Qualcomm Snapdragon X24 & Android 12 & G020J & 2019 \\
\textbf{OnePlus 7 Pro} & Qualcomm Snapdragon 855 & Android 11 & GM1910 & 2019 \\
\rowcolor{gray!25}
\textbf{Google Pixel 3a} &  Qualcomm Snapdragon 670  & Android 11 & G020F & 2019 \\
\textbf{Samsung Note 10 5G} & Samsung Exynos 9825 & Android 10 & SM-N976Q & 2018 \\
\rowcolor{gray!25}
\textbf{Huawei P8 Lite 2017} & HiSilicon Kirin 655 & Android 10 & PRA-LA1 & 2017 \\
\textbf{Apple iPhone 7} & Intel XMM7360 & iOS 15 & A1778 & 2016 \\
\rowcolor{gray!25}
\textbf{Apple iPhone 5} & Qualcomm MDM9615M & iOS 10 & A1428 & 2013 \\
\bottomrule
\end{tabular}
}
\end{table*}

\begin{table*}[bth]
\centering
\small
\caption{\label{tab:counts} Number of SMS received in our experiments. The $^\ast$ denotes a sender device only.}

\begin{tabular}{lcccccccccccc}
\toprule
Device& \multicolumn{12}{c}{Countries and Operators}\\
\midrule

\emph{Int'l, Nat. \& Reg.}&\multicolumn{3}{c}{AE} & \multicolumn{2}{c}{GR} & UK & US & DE & DK\\
\emph{(Sections~\ref{sec:classification-intl} \&~\ref{national-regional})} & A & B & C & C & D & H & J & E & I\\
\cmidrule{2-10}

\textbf{Apple iPhone 13} & 0 & 0 & 0 & 350 & 0 & 0 & 0 & 0 & 0\\
\rowcolor{gray!25}
\textbf{One Plus Nord 2 5G} & 350 & 350 & 0 & 0 & 0 & 0 & 0 & 0 & 0\\
\textbf{Alcatel 1S} & 0 & 0 & 0 & 350 & 0 & 0 & 0 & 0 & 0\\
\rowcolor{gray!25}
\textbf{Apple iPhone 12 mini} & 350 & 350 & 0 & 0 & 0 & 0 & 0 & 0 & 0\\ 
\textbf{Nokia 8.3 5G} & 350 & 350 & 0 & 0 & 0 & 0 & 0 & 0 & 0\\
\rowcolor{gray!25}
\textbf{Apple iPhone 12} & 0 & 0 & 0 & 350 & 350 & 350 & 0 & 0 & 0\\
\textbf{Samsung Galaxy A21S} & 0 & 0 & 0 & 350 & 0 & 0 & 0 & 0 & 0\\
\rowcolor{gray!25}
\textbf{Huawei P40 Pro 5G$^\ast$} & 0 & 0 & 0 & 0 & 0 & 0 & 0 & 0 & 0\\ 
\textbf{Google Pixel 4 XL} & 0 & 0 & 0 & 0 & 0 & 0 & 350 & 0 & 0\\
\rowcolor{gray!25}
\textbf{Samsung Note 10 5G} & 350 & 350 & 0 & 0 & 0 & 0 & 0 & 0 & 0\\
\textbf{Apple iPhone 7} & 0 & 0 & 0 & 700 & 0 & 0 & 0 & 0 & 350\\
\rowcolor{gray!25}
\textbf{Apple iPhone 5} & 0 & 0 & 350 & 0 & 0 & 0 & 0 & 0 & 0 \\
\textbf{OnePlus 7 Pro} & 0 & 0 & 0 & 0 & 0 & 0 & 0 & 350 & 0 \\

\cmidrule{2-13}
 \emph{Nat. \& Reg.} & \multicolumn{3}{c}{BE} & \multicolumn{3}{c}{DE} & \multicolumn{3}{c}{LU} & \multicolumn{3}{c}{NL} \\ 
\emph{(Section~\ref{national-regional})} & E & F & G & E & F & G & E & F & G & E & F & G \\
\cmidrule{2-13}
\rowcolor{gray!25} \textbf{Nokia 5.3} & 0 & 0 & 798 & 1350 & 0 & 3159 & 0 & 0 & 455 & 0 & 0 & 1419\\
\textbf{OnePlus 7 Pro} & 0 & 0 & 839 & 2021 & 0 & 3109 & 0 & 0 & 422 & 0 & 0 & 1411 \\
\rowcolor{gray!25}\textbf{Google Pixel 3a} & 0 & 818 & 0 & 1963 & 2516 & 1092 & 0 & 499 & 0 & 0 & 1399 & 0\\
\textbf{Huawei P8 Lite 2017} & 804 & 0 & 0 & 3342 & 1111 & 1153 & 513 & 0 & 0 & 1416 & 0 & 0 \\

\cmidrule{2-13}
\emph{Temporal \& Network}&&&& \multicolumn{3}{c}{DE} &&&& \multicolumn{3}{c}{NL} \\ 
\emph{(Sections~\ref{sec:temporal-stability} \&~\ref{sec:network-analysis})} &&&& E & F & G &&&& E & F & G \\
\cmidrule{2-13}

\rowcolor{gray!25}\textbf{Huawei P8 Lite 2017} &&&& 14132 & 0 & 0 &&&& 15607 & 0 & 0 \\
\textbf{OnePlus 7 Pro} &&&& 0 & 0 & 16625 &&&& 0 & 0 & 12799 \\
\rowcolor{gray!25}\textbf{Google Pixel 6a} &&&& 0 & 14752 & 0 & & & & 0 & 16115 & 0 \\
\textbf{Samsung Galaxy A53} &&&& 0 & 0 & 11095 & & & & 0 & 0 & 15778 \\

\end{tabular}

\end{table*}


\end{document}